# Pairing particles into holonomies

Vera Neef, Matthias Heinrich, Tom A.W. Wolterink, Alexander Szameit*

Institute for Physics, University of Rostock; Albert-Einstein-Str. 23, 18059 Rostock, Germany

* alexander.szameit@uni-rostock.de

**Abstract:**

Holonomies are of great interest to quantum computation and simulation. The geometrical nature of these entities offers increased stability to quantum gates. Furthermore, symmetries of particle physics are naturally reflected in holonomies, making them ideally suited for quantum simulation of quantum chromodynamics and grand unified theories. Yet, practically designing quantum holonomies with the required properties and scale is challenging. Here, we construct a new class of holonomies by increasing the particle number. We show that multi-particle holonomies can even exist in systems devoid of any single-particle holonomies. We present a comprehensive framework for multi-particle quantum holonomies and experimentally realize various two-particle holonomies in integrated photonics. Our results enable particle number to be harnessed as a design parameter, offering increased freedom in constructing holonomies for quantum computation and simulation.

Holonomies, geometrical transformations arising from the presence of a gauge field, are deeply intertwined with many of the core topics in modern physics, ranging from quantum computing to topology, field theories and grand unified theories. Gauge fields naturally emerge in various theoretical frameworks. Their perhaps most familiar example manifests as a scalar field that gives rise to electric and magnetic fields. In turn, non-Abelian gauge theories have unified electromagnetism with weak and strong interactions (*1*), inspiring modern frameworks such as lattice gauge theory (*2*, *3*), topological quantum field theory (*4*, *5*), and loop quantum gravity (*6*). In each case, gauge fields give rise to geometric phases and, in particular, holonomies. A quintessential example of a geometric phase is the Aharonov-Bohm effect (*7*), where an electron encircling a solenoid accumulates a phase proportional to the magnetic flux, arising purely from geometry without direct interaction with any observable field.

Interestingly, holonomies can also be intentionally created in systems without inherent gauge fields by constructing cyclic subspaces (*8*–*11*). Within such a subspace, the evolution of a quantum state separates into two contributions: a dynamical component, representing the system's internal clock, and a geometric component governed by a gauge field that does not vanish under any gauge transformation. This latter geometric aspect becomes a defining feature of the system (*8*–*12*). Moreover, when the dynamical contribution vanishes, the resulting evolution is called a quantum holonomy, characterized by its purely geometric nature, offering exceptional stability (*13*–*17*). This robustness has practical implications, particularly in holonomic quantum computing (*18*–*20*). Quantum holonomies also hold promise for simulating fundamental physics by studying systems that transform under shared symmetries. For instance, spin and weak or strong isospin obey similar symmetries, meaning insights from one phenomenon can inform our understanding of the others. Similarly, studying quantum systems that generate specific holonomies can enhance our grasp of elementary particles like gluons and quarks, which remain experimentally elusive (*21*). Last but not least, holonomies have also become pivotal in exploring topological phases of matter (*22*–*25*).

Yet, despite their potential, realizing quantum systems that produce holonomies with specific properties and symmetries remains a significant challenge. Previous research has predominantly focused on single-particle systems, with experiments conducted across platforms (*20*, *26*) including cold atoms (*27*), trapped ions (*28*, *29*), superconductors (*30*), nuclear magnetic resonance (*31*), solid-state spins (*32*), and photonics (*33*–*35*). However, advancing the concept of quantum holonomies to encompass multi-particle quantum features is crucial for designing simulations capable of capturing complex fundamental physics phenomena. To date, only one experimental study has demonstrated a specific two-particle holonomy (*26*, *36*), and even theoretical work on multi-particle holonomies remains limited (*37*–*40*).

In this study, we pioneer the exploration of multi-particle quantum holonomies both theoretically and experimentally. We develop a comprehensive framework for quantum holonomies in multi-particle systems. In taking a particularly close look at two-particle systems, we demonstrate how particle number can serve as a tool to design holonomies and expand the parameters available for creating holonomies with tailored properties. Remarkably, introducing a second particle unveils entirely new holonomies. Moreover, two-particle holonomies can arise even in systems where single-particle holonomies are absent. To validate our theoretical findings, we experimentally realize a plethora of two-particle holonomies using a photonic platform as a proof of concept. Our work significantly broadens the scope of quantum holonomies, opening new avenues for designing quantum simulations to probe fundamental physics phenomena.

## Holonomies

We define quantum holonomies as purely geometric transformations within cyclic subspaces. To clarify this definition, let us break it down step by step.

*Spaces and subspaces:* In quantum mechanics, the Hilbert space encompasses all possible states of a given system. For example, in Fig. 1, the pink layer represents a system where a single particle can occupy four distinct modes. These four basis states form the Hilbert space, which includes all possible superpositions of these states. A subspace of the Hilbert space contains only a subset of these states, thereby reducing the number of basis states. For instance, the pale pink domain surrounded by a dashed border in Fig. 1 represents a subspace that includes only two of the four basis states. Subspaces can be defined not only by restricting certain modes, but also by specifying particle numbers when multiple particles are involved. For example, the pale pink subspace in Fig. 1 is defined by restricting the particle to specific modes. If a second particle is introduced into this subspace, its size increases. For example, with two indistinguishable bosons confined to two modes, there are three possible configurations, forming a subspace with three basis states, illustrated by the orange layer in Fig. 1. Additionally, we can define a sub-subspace within this space, such as the pale orange region, which encompasses only states where both bosons occupy the same mode. Thus, increasing the particle number introduces an additional degree of freedom for defining subspaces.

*Cyclic Subspaces:* Consider a quantum state evolving over time. As it evolves, it appears different at each moment. Similarly, if we examine the basis states forming a subspace, the entire subspace evolves and becomes time-dependent. A subspace is deemed cyclic if, after some evolution, it returns to its initial configuration at the endpoint in time. Importantly, this does not require each individual state within the subspace to return to its original state—only the subspace as a whole must be recovered.

*Dynamical and geometric contributions:* The evolution of a quantum system is governed by the time-dependent Hamiltonian, $\hat{H} = \hat{H}(t)$, if its full Hilbert space is considered. However, within a cyclic subspace, the evolution consists of two distinct contributions: dynamical and geometric. The dynamical contribution is described by an operator $\hat{K}$, which resembles the Hamiltonian within the subspace. Meanwhile, the geometric contribution arises from the time-dependence of the subspace itself, governed by the gauge field $\hat{A}$. This geometric contribution is not directly dependent on the system's Hamiltonian and cannot be eliminated through any gauge transformation. The time-evolution operator in a cyclic subspace is given by (*11*):

$$\hat{\mathcal{U}}(T) = \mathcal{P} \exp \int_0^T \mathrm{i} \left(\hat{A} - \hat{K}\right) \mathrm{d}t \,, \tag{1}$$

where $T$ is the endpoint in time, $\mathcal{P}$ denotes path ordering, and i is the imaginary unit. When the dynamical contribution vanishes ($\hat{K} = 0$), the evolution becomes holonomic, and $\hat{\mathcal{U}}(T)$ is the holonomy – a purely geometric transformation within a cyclic subspace.

## Creating multi-particle holonomies

Let us now explore how increasing the particle number can lead to the creation of new holonomies. Previous studies have demonstrated that a higher particle number can enlarge the size of an existing holonomy (*36*, *38*). In the following, we will show that entirely new holonomies can also emerge when the particle number increases. The methods section contains a more general discussion of linear and Hermitian systems with multiple particles. Here, we will consider the special case of a two-particle system described by the coupled-mode Hamiltonian:

$$\hat{H} = \sum_{k,j} \kappa_{jk}\, \hat{a}_k^\dagger \hat{a}_j, \tag{2}$$

where $\hat{a}_k^\dagger$ and $\hat{a}_j$ are the creation and annihilation operators, which can be either bosonic or fermionic. The coupling coefficients $\kappa_{jk}$ represent the interaction (or hopping) between the $k$-th and $j$-th modes, and hermiticity demands $\kappa_{jk} = \kappa_{kj}^*$ (where * denotes the complex conjugate). The terms $\kappa_{jj}$ represent self-coupling and mode detuning. To describe the basis states of cyclic subspaces, we introduce (orthonormal) mode operators $\hat{\psi}_k^\dagger = \hat{\psi}_k^\dagger(t)$, which evolve according to the Heisenberg equation of motion and are general superpositions of $\hat{a}_j^\dagger$. The basis states of a single-particle cyclic subspace are represented as $\hat{\psi}_k^\dagger|0\rangle$, and for a two-particle subspace, as $\hat{\psi}_k^\dagger\hat{\psi}_j^\dagger|0\rangle$ (up to normalization). Using these operators, we define $\mathcal{J}_{jk} = \langle 0|\hat{\psi}_j \hat{H} \hat{\psi}_k^\dagger|0\rangle$, which we refer to as the coupling between the modes $\hat{\psi}_k^\dagger$ and $\hat{\psi}_j^\dagger$. While this interpretation is strictly accurate only when the Hamiltonian commutes with the time-evolution operator, it is a useful simplification for our discussion.

For a holonomy to form, the dynamical contribution to the evolution must vanish ($\hat{K} = 0$). In a single-particle system, this requires all couplings $\mathcal{J}_{jk}$ within the cyclic subspace to vanish. Importantly, this does not mean every input state leaves the system unaltered; couplings outside the cyclic subspace can influence the evolution via the gauge field $\hat{A}$. In two-particle systems, however, the holonomic conditions are significantly less restrictive. This is the central result of our work, illustrated by the two-particle holonomic conditions in Fig. 2. To construct a two-particle holonomy:

1. Start with a cyclic subspace and determine the particle type (bosons, fermions, or distinguishable particles), as the holonomic conditions differ for each case.
2. Check the conditions for individual basis states, shown in the upper orange section of Fig. 2.
    - Basis states where both particles occupy the same mode are subject to different conditions than those where the particles occupy different modes.
    - For fermions, states where two particles occupy the same mode do not exist, so the corresponding cells are marked as invalid.
3. Verify conditions for pairs of basis states, depicted in the pink section of Fig. 2.
    - Each pair of states must satisfy a condition ensuring that one matrix element of $\hat{K}$ vanishes.
    - In some cases, marked with ticks in Fig. 2, no additional conditions are required for the element to vanish.

Interestingly, even if a condition requires a coupling $\mathcal{J}_{jk}$ to vanish, it pertains only to the coupling between two specific modes, regardless of how many modes contribute to the two states in question.

Figure 2 highlights the greater design flexibility of two-particle holonomies compared to single-particle ones, where all couplings within the subspace must vanish. Note that, two-particle holonomies can arise even if not all couplings $\mathcal{J}_{jk}$ vanish, forming a new class of holonomies. This newly discovered freedom in creating holonomies is increased even more for higher particle numbers (see methods) and opens up a wealth of possibilities. To illustrate this, in the next section, we examine a small example system and implement it experimentally using a photonic platform. By measuring holonomic signatures across multiple cyclic subspaces, we demonstrate the creativity and versatility involved in finding two-particle cyclic subspaces with vanishing dynamical contributions.

## Experimental implementation of two-photon holonomies

As an example, we investigate a planar structure with four waveguides, as shown in Fig. 3a. Each waveguide supports a single mode, and we construct our basis such that the propagating modes $\hat{\psi}_k^\dagger$ coincide with the waveguide modes at the input. Since holonomies are associated with stability (*13–17*), we test how a transformation behaves under distortions.

Seven variations of the waveguide structure with different propagation lengths are fabricated using the femtosecond laser-direct-writing technique (*41*). The couplings between modes are finely tuned to ensure that, at the ideal length, the structure performs two independent flip operations: one between the first and fourth modes and another between the second and third modes (Fig. 3a). However, for lengths longer or shorter than the ideal, the cyclicity condition is violated, resulting in arbitrary transformations. To satisfy the conditions outlined in the orange section of Fig. 2, detuning between all modes was set to zero. Probing this system with pairs of photons produces numerous cyclic subspaces, making this waveguide structure an ideal system for studying holonomies. For single photons, the full Hilbert space is four-dimensional, corresponding to the four modes. This allows the definition of 14 distinct subspaces (excluding the full Hilbert space), of which only two are cyclic. When two indistinguishable photons are considered, the Hilbert space expands to 10 dimensions, representing the 10 ways to distribute two photons across four modes. This configuration generates 1,022 possible subspaces, 62 of which are cyclic due to the symmetry of the system. We experimentally implement all cyclic subspaces and compare subspaces that give rise to holonomies with those that do not.

To examine a given cyclic subspace, it must first be experimentally enclosed, as the distortions break the cyclicity. Consider the subspace where a single photon is launched into either the first or fourth mode. The experimental enclosure involves two steps: (i) Input initialization, where the photon is launched into the initial subspace, either into the first mode (in one experiment) or the fourth mode (in another), and (ii) output post-selection. For a photon launched into the first mode, the ideal flip operation predicts its detection in the fourth mode. Any detection in the fourth mode is counted as a success, while detection in the first mode is counted as a failure. Detections in the second or third modes are disregarded as they fall outside the cyclic subspace. Similarly, for photons launched into the fourth mode, detections in the first mode are counted as successes. Since there is no direct coupling between the first and fourth modes, the transformation within this subspace is expected to be holonomic and thus stable against distortions. This stability is evident in a plateau of success probability as a function of propagation length, shown in Fig. 3b. The plateau has a width of approximately 13 mm, indicating a single-photon holonomy. Launching two indistinguishable photons into the same two modes also creates a holonomy (*38*). This two-photon subspace comprises three basis states (Fig. 3c). Using a similar measurement scheme with photon-number-resolving detection (see methods), the transformation again demonstrates stability, with a plateau width of about 13 mm. Further subdivision of this subspace into a sub-subspace comprising only the two bunched states (where both photons occupy the same mode) reveals a holonomy with even greater stability, with a plateau width of approximately 15 mm (Fig. 3d).

The second cyclic single-photon subspace, where the photon is launched into either the second or third mode, behaves differently (Fig. 3e). Post-selection is applied to detections in these two modes, but this subspace does not give rise to a holonomy. The transformation has a dynamical contribution governed by the non-vanishing coupling between the second and third modes, leading to poor stability. The detection probability shows a plateau width of only about 3 mm. Injecting two indistinguishable photons into these modes (Fig. 3f) similarly fails to create a holonomy. As per the fifth pink condition in Fig. 2, the coupling between the second and third modes governs the dynamical contribution, resulting in a narrow plateau of about 1 mm. However, defining a sub-subspace that includes only the two bunched states (where both photons occupy the same mode) removes the influence of this coupling. In this configuration, the last pink condition in Fig. 2 is the only relevant pink condition which always ensures a vanishing dynamical contribution without imposing restrictions on the coupling. As a result, this sub-subspace produces a holonomy, with improved stability reflected in a plateau width of about 10 mm (Fig. 3g).

This highly symmetric waveguide structure, with its abundance of cyclic subspaces, supports 18 distinct non-Abelian holonomies for two indistinguishable photons (see supplementary text). Generally, subspaces with fewer basis states are more likely to exhibit holonomies, while larger subspaces typically do not. However, exceptions exist. For example, Fig. 3f shows a rare non-holonomic subspace with three basis states. The largest holonomic subspace in this system with six basis states demonstrates that creating large holonomies becomes more feasible with more particles.

One unexplored parameter is the particle type. As indicated by the conditions in Fig. 2, the type of particle influences whether a holonomy arises. In the next section, we explore a specific subspace that fails to produce a holonomy with two indistinguishable photons but does so when the particles are distinguishable.

Consider the subspace illustrated in Fig. 4a, which includes two indistinguishable photons distributed across four input states. Due to the second pink condition in Fig. 2, the dynamical contribution in this case is governed by the coupling between the second and third modes. As a result, the transformation is non-holonomic, with a limited stability reflected in a plateau width of approximately 3 mm. Now, let us replace the indistinguishable photons with two distinguishable photons in the same subspace. In this configuration, there are eight possible input states. However, since the states represented on the left-hand and right-hand sides of the legend produce identical results, Fig. 4b shows only four curves for simplicity, while all eight states are considered on the output side. If any input state from the left-hand side is prepared, all states from the right-hand side are treated as failures. Consistent with the non-holonomic nature of this transformation, the plateau width remains around 3 mm. This cyclic subspace, however, can be further subdivided into two identical subspaces, each with only four input states, as shown in Fig. 4c. The experimental enclosure of this sub-subspace exploits the distinguishability of the particles. Specifically, if photon $a$ is detected in the mode where photon $b$ was expected (and vice versa), this event is no longer recorded as a failure (as in Fig. 4b). Instead, it is disregarded, as it lies outside the defined sub-subspace. In this refined sub-subspace, the dynamical contribution vanishes because the second pink condition from Fig. 2 no longer applies. Instead, the third pink condition guarantees a vanishing dynamical contribution, regardless of specific couplings. The resulting transformation is purely holonomic, exhibiting enhanced stability with a plateau width of approximately 10 mm. This example underscores the critical influence of particle type in constructing holonomies. The ability to tailor the dynamical contribution by considering particle properties expands the design possibilities for holonomic transformations.

## Conclusion

In this work, we introduced a general condition for multi-particle holonomies in linear and Hermitian quantum systems that ensures vanishing of the dynamical contribution in any cyclic multi-particle subspace. In particular, we discover a rich variety of two-particle holonomies, many of which appear even when no single-particle holonomies exist. In a proof-of-principle experimental system, where a single photon generates only a single holonomy spanning two basis states, we observe 18 distinct non-Abelian holonomies with two indistinguishable photons, spanning between two and six basis states. While the holonomic conditions for two indistinguishable bosons and fermions are generally the same, distinguishable particles can behave differently, leading to holonomies that are unique to either indistinguishable or distinguishable particles. We experimentally demonstrate such an instance, extending the concept of multi-particle holonomies to classical experiments.

Previous studies on multi-particle holonomies focused on enlarging a single-particle holonomy by introducing more particles addressing the dynamical contribution at the mode level, rather than the state level, hence using a Heisenberg picture (*36*, *38*, *40*). Ensuring the dynamical contribution vanishes at the mode level (i.e., all $\mathcal{J}_{jk} = 0$) reliably produces holonomies for any particle number, and serves as a robust tool for confirming the existence of holonomies across all particle numbers, including single-particle systems. However, by addressing the dynamical contribution at the state level and using a Schrödinger picture, as demonstrated in this work, we uncover a wealth of additional multi-particle holonomies that may emerge in larger subspaces. Furthermore, many cyclic subspaces would be missed if defined at the mode level rather than the state level. Looking forward, we hope to extend these methods to non-linear and non-Hermitian systems.

This work sheds new light on the creation of holonomies involving multiple particles. The flexibility in designing multi-particle holonomies, which imposes fewer restrictions on system couplings, could play a significant role in developing complex holonomies necessary for simulating quantum systems, such as lattice gauge theory, topological quantum field theory, loop quantum gravity, and quantum chromodynamics.

**Acknowledgments:** The authors thank Karo Becker, Friederike Klauck and Òscar Perearnau Herrero for many fruitful discussions, Toni Gerdes for help with several combinatorics problems and C. Otto for preparing the high-quality fused silica samples used for the inscription of all photonic structures employed in this work, as well as Franklin Martinez and Jelena Boldt for all their help.

**Funding:** A.S. acknowledges funding from the Deutsche Forschungsgemeinschaft (grants SZ 276/9-2, SZ 276/19-1, SZ 276/20-1, SZ 276/21-1, SZ 276/27-1), as well as the Krupp von Bohlen and Halbach Foundation. A.S. also acknowledges funding from the FET Open Grant EPIQUS (grant no. 899368) within the framework of the European H2020 programme for Excellent Science. T.A.W.W. is supported by a Marie Skłodowska-Curie fellowship from the European Commission (project no. 895254). A.S. acknowledge funding from the Deutsche Forschungsgemeinschaft via GRK 2676/1-2023 'Imaging of Quantum Systems' (project no. 437567992). A.S. and M.H. acknowledge funding from the Deutsche Forschungsgemeinschaft via SFB 1477 'Light–Matter Interactions at Interfaces' (project no. 441234705).

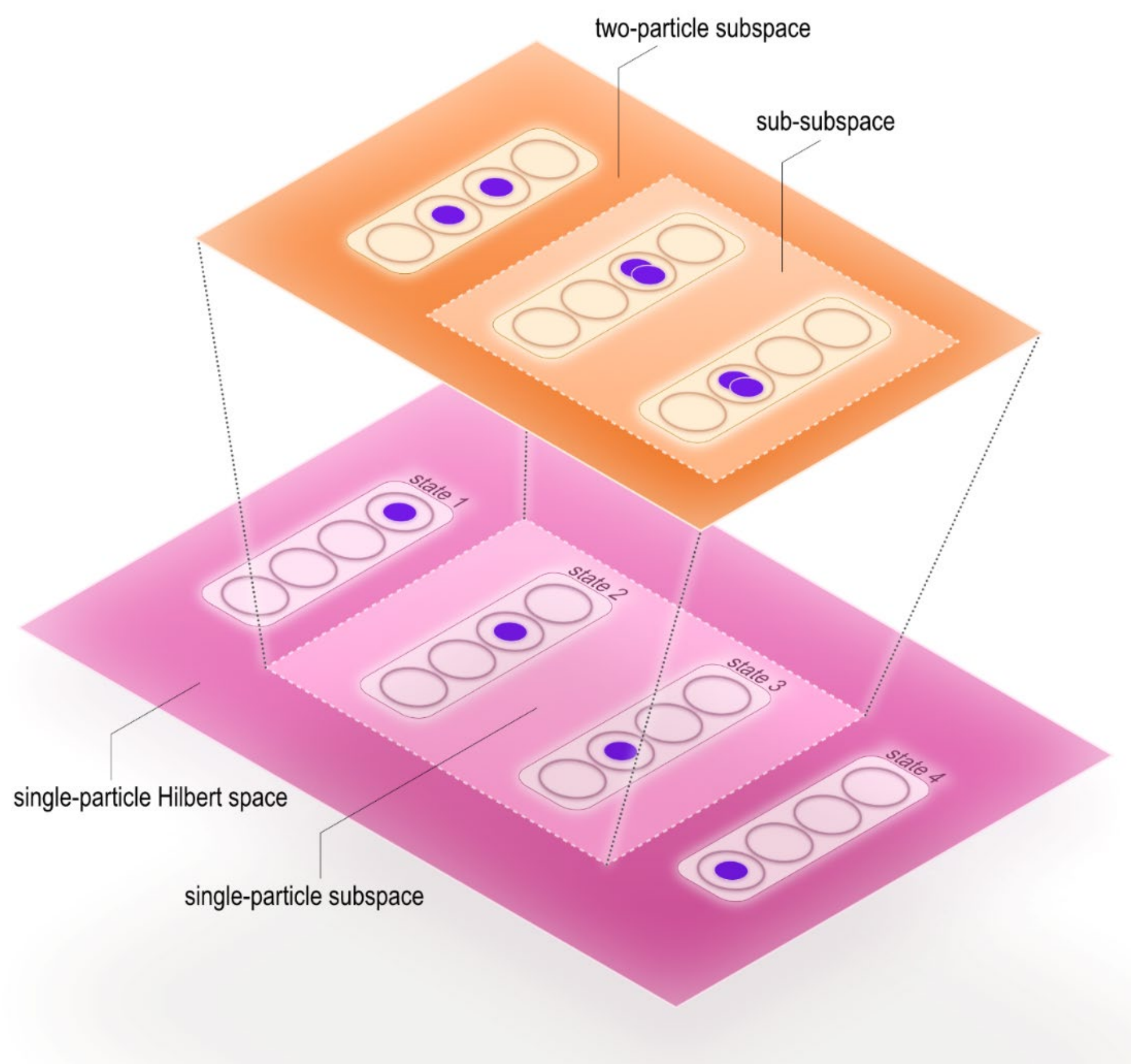

**Figure 1: Illustration of the concept of subspaces.** Consider a system with four modes, where the full Hilbert space is spanned by four basis states, as shown in the bright pink layer at the bottom. A subspace contains only a subset of these basis states. For example, the pale pink section includes just two of the four basis states. When two particles are distributed across the modes, the number of possible states and subspaces increases. For instance, there are three ways to distribute two indistinguishable bosons between modes two and three. The bright orange layer represents a subspace of the two-boson Hilbert space, which spans 10 basis states. This two-boson subspace can be further subdivided: the pale orange section represents a sub-subspace containing only the two basis states in which both bosons occupy the same mode. This demonstrates the added flexibility in defining subspaces with two particles, compared to a single particle.

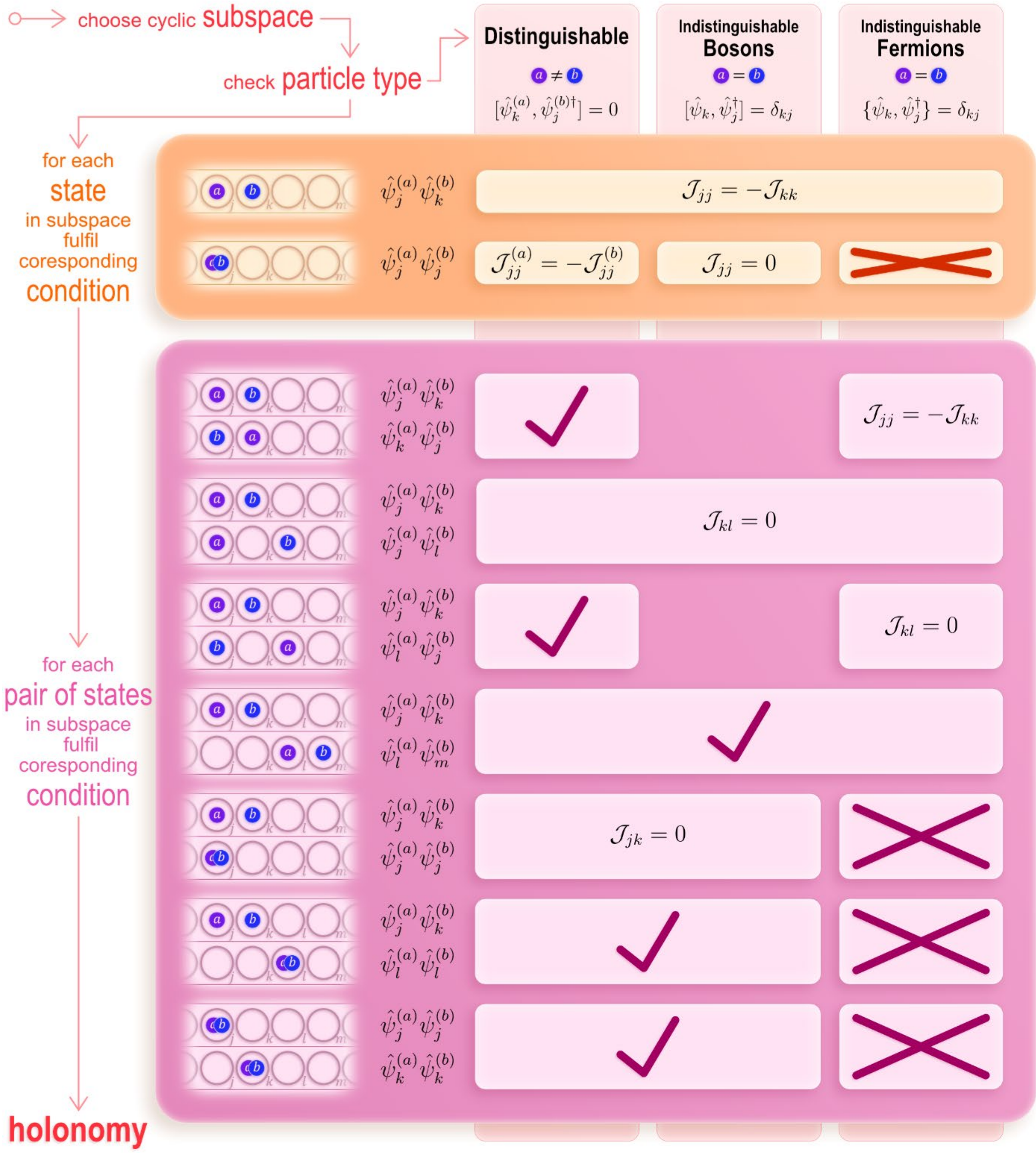

**Figure 2: Two-particle holonomic conditions.** Consider a linear and Hermitian system that is governed by a coupled mode Hamiltonian with (orthonormal) mode operators $\hat{\psi}_k^\dagger$ that evolve according to the Heisenberg equation of motion. The states $\hat{\psi}_k^\dagger \hat{\psi}_j^\dagger |0\rangle$ (up to normalization) form a basis for a cyclic subspace. To ensure that the dynamical contributions vanish and a holonomy is created, the following conditions must be met: 1) Particle type: Begin by identifying the appropriate column of conditions based on the particle type. 2) State conditions: For each basis state of the cyclic subspace, satisfy the corresponding condition in the upper orange section. Check whether the particles occupy the same mode or different modes. 3) Pairing conditions: For each pair of basis states in the cyclic subspace, verify the corresponding condition. Consider all possible pairs and ensure the correct identification of whether the states occupy the same or different modes. The conditions on the right-hand side have the following meanings: (i) If the field refers to two fermions occupying the same mode, it is crossed out due to the unphysical

nature of such states. (ii) A tick mark indicates no further conditions are needed for that state pair, as the dynamical contribution vanishes due to mode orthogonality. (iii) A missing field indicates redundancy in the condition. (iv) $\mathcal{J}_{jk} = \langle 0 | \hat{\psi}_j \hat{H} \hat{\psi}_k^\dagger | 0 \rangle$, with $\hat{H}$ being the Hamiltonian. Thus, the orange section can be interpreted as conditions for detuning specific modes, and the pink section refers to conditions for couplings between modes. If all conditions are satisfied for all states and pairs within the subspace, a holonomy is formed.

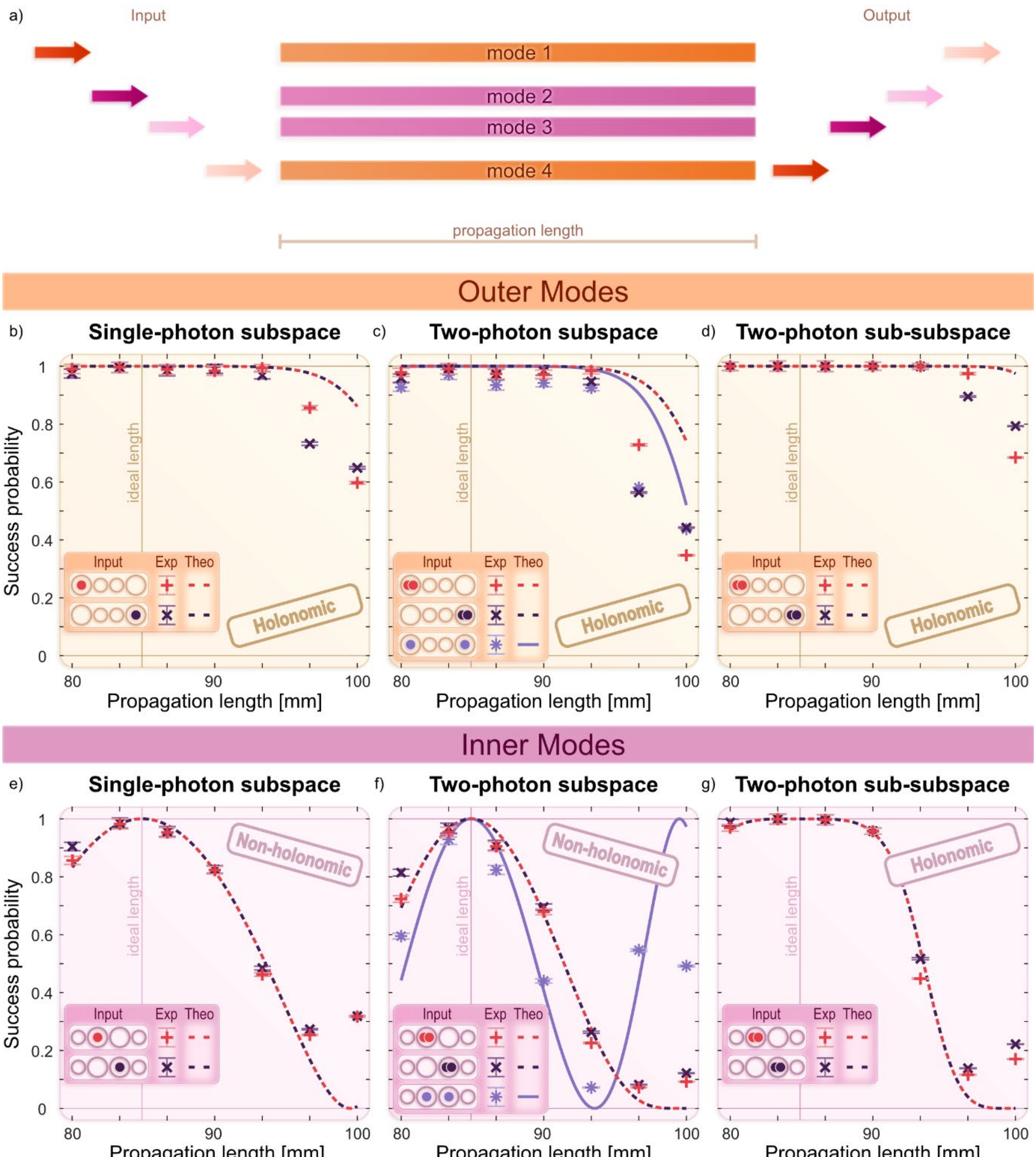

**Figure 3: Examples of single-photon and two-photon holonomies in a specific four-waveguide structure.** (a) Sketch of a four-mode structure: The structure comprises four non-detuned modes arranged in a planar configuration with a specific coupling distribution. At an ideal length of 84.9 mm, the structure performs two flip operations: light launched into the first mode exits via the fourth mode, and vice versa (orange arrows). Similarly, light launched into the second mode exits via the third mode, and vice versa (pink arrows). (b)–(g) Success probability over propagation length in different cyclic subspaces: The propagation length of the four-waveguide structure is varied. For each input state, the probability of detecting the expected output state when post-selecting onto the cyclic subspace is measured. Each subspace is unambiguously characterized by specific input states: (b) Outer mode subspace (single photon): A photon is launched into either the first or fourth mode. This subspace exhibits holonomic properties, showing enhanced stability to propagation length variations with an experimental plateau width of approximately 13 mm. (c) Outer mode subspace (two photons): Two photons are launched into the outer modes. The experimental plateau width of ~13 mm indicates holonomic features. (d) Restricted subspace (outer modes): When both photons in all input states occupy the same mode, the plateau widens to ~15 mm. (e) Inner mode subspace (single photon): A single photon launched into either of the inner modes results in non-holonomic transformations, with poor stability against propagation length changes (plateau width ~3 mm). (f) Inner mode subspace (two photons): Launching two indistinguishable photons into the inner modes also yields a non-holonomic transformation, with a plateau width of ~1 mm. (g) Restricted subspace (outer modes): By further restricting the inner subspace to only cases where both photons occupy the same mode, a holonomic transformation emerges. This is indicated by increased stability and a plateau width of ~10 mm.

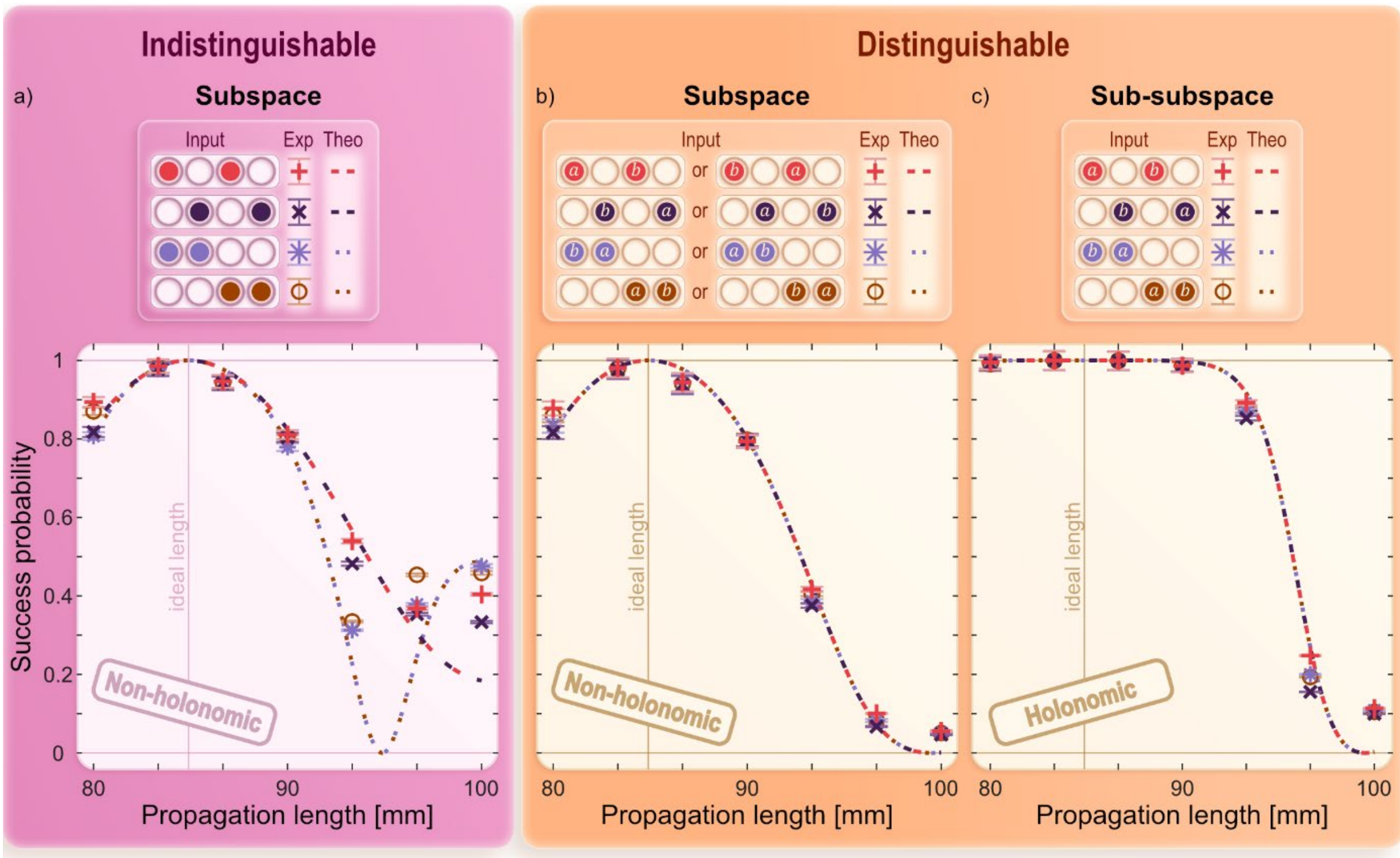

**Figure 4: Stability of cyclic subspaces with indistinguishable and distinguishable pairs of photons.** (a) Non-holonomic cyclic subspace (two indistinguishable photons): Two indistinguishable photons are launched into a cyclic subspace with non-vanishing dynamical contribution to the evolution. All four input states exhibit poor stability of the success probability against changes in propagation length, with an average experimental plateau width of approximately 3 mm. (b) Non-holonomic cyclic subspace (two distinguishable photons): When two distinguishable photons are launched into the same cyclic subspace, its size increases to eight basis states. As the input states on the left side of the legend yield identical results to those on the right side, only four theoretical curves and four corresponding measurement datasets are shown. Despite this expansion, the non-vanishing dynamical contribution makes the evolution non-holonomic, resulting in poor stability with a plateau width of approximately 3 mm. (c) Sub-divided subspace (holonomic): The subspace from (b) can be further divided into two identical sub-subspaces. By restricting to a sub-subspace containing only four basis states, the dynamical contribution is eliminated, yielding a purely holonomic transformation. This results in significantly enhanced stability against propagation length variations, with an experimental plateau width of approximately 10 mm.

# Supplementary Materials for

## Pairing particles into holonomies

Vera Neef, Matthias Heinrich, Tom A.W. Wolterink, Alexander Szameit*

Corresponding author: *alexander.szameit@uni-rostock.de

## Materials and Methods

Mathematical methods

The four-waveguide system discussed in the main text is described by the Hamiltonian

$$\hat{H}(t) = \Omega(t)\left(\frac{\sqrt{3}}{2}\hat{a}_1^\dagger\hat{a}_2 + \hat{a}_2^\dagger\hat{a}_3 + \frac{\sqrt{3}}{2}\hat{a}_3^\dagger\hat{a}_4\right) + \mathrm{h.c.}$$

where $\hat{a}_k^{(\dagger)}$ $(k = 1,2,3,4)$ are the bosonic annihilation (creation) operators in the waveguide modes. The time-dependency of the couplings is moved to a global (real and positive) envelope function $\Omega(t)$, ensuring that the Hamiltonians at different times commute $\hat{H}(t_1)\hat{H}(t_2) = \hat{H}(t_2)\hat{H}(t_1)$. Consequently, the time-evolution operator (for the creation operators) can be evaluated explicitly as $\hat{U}(t) = \exp(-\mathrm{i}\int_0^t \hat{H}(\tau)\mathrm{d}\tau)$. The weights of the couplings (1 and $2/\sqrt{3}$) are constant and ensure its cyclic behavior, effectively realizing a four-waveguide Jx-lattice (*1*). Defining $\delta(t) = \int_0^t \Omega(\tau)\mathrm{d}\tau$, after a time T (referred to in the main text as ideal length) with $\delta(T) = \pi$ one cycle is completed and the structure has completed the double-flip operation

$$\hat{U}(T) = \mathrm{i}\begin{pmatrix} 0 & 0 & 0 & 1 \\ 0 & 0 & 1 & 0 \\ 0 & 1 & 0 & 0 \\ 1 & 0 & 0 & 0 \end{pmatrix}.$$

Here, similar to a beam-splitter matrix, a matrix notation with regard to $\hat{a}_k^\dagger$ is used, i.e.

$$\begin{pmatrix} \hat{a}_1^\dagger(T) \\ \hat{a}_2^\dagger(T) \\ \hat{a}_3^\dagger(T) \\ \hat{a}_4^\dagger(T) \end{pmatrix} = \hat{U}(T)\begin{pmatrix} \hat{a}_1^\dagger(0) \\ \hat{a}_2^\dagger(0) \\ \hat{a}_3^\dagger(0) \\ \hat{a}_4^\dagger(0) \end{pmatrix}.$$

In a cyclic subspace, the time-evolution reads $\hat{\mathcal{U}}(T) = \mathcal{P}\exp\int_0^T \mathrm{i}\left(\hat{A} - \hat{K}\right)dt$ (*2*). To calculate $\hat{A}$, known as gauge field or connection, the usual method is to find a cyclic basis $|\Phi_k(0)\rangle = |\Phi_k(T)\rangle$ for the subspace. Then its elements read $A_{kj} = \mathrm{i}\langle\Phi_k|\mathrm{d}_t|\Phi_j\rangle$. This non-adiabatic and non-Abelian approach to holonomies, first described by Anandan (*2*), is used as it is a generalization to several (more widely known) approaches: For adiabatic and non-Abelian Wilczek-Zee holonomies (*3*), the states $|\Phi_k\rangle$ are degenerate and time-dependent eigenstates. The changes have to occur slowly to ensure adiabaticity, i.e. avoid crossing to other energy-levels. The well-known Berry phase (*4*) is adiabatic and Abelian. Here, only a single eigenstate is concerned and the holonomy becomes a scalar phase contribution. For the non-adiabatic and Abelian Aharonov-Anandan phase (*5*), the cyclic basis shrinks to only one state $|\Phi(0)\rangle = |\Phi(T)\rangle$, that is not required to be an eigenstate and the holonomy shrinks to a scalar phase factor. While the theory of two-particle holonomies is valid for any of the afore mentioned types of holonomies, the specific holonomies discussed in this work fall into the category of non-adiabatic and non-Abelian holonomies as described by Ananadan (*2*). Since the cyclic subspace is not spanned by eigenstates, the holonomies are considered non-adiabatic even though, due to its construction, the eigenstates of the Hamiltonian are time-independent.

Expanding on the standard method, to simplify the calculations, we propose a non-cyclic basis that preserves the geometric properties. Defining the basis modes as $\hat{\phi}_k^\dagger(t) = \exp(-\mathrm{i}\,\delta(t)/2)\,\hat{a}_k^\dagger(t)$ ($k = 1,2,3,4$), where $\hat{a}_k^\dagger(t)$ propagates according to the time-evolution operator $\hat{U}(t)$, this basis exploits the highly symmetric nature of this specific Hamiltonian and allows for a calculation of all holonomies using the same basis. Since $\hat{\phi}_1^\dagger(T) = \hat{\phi}_4^\dagger(0)$, $\hat{\phi}_2^\dagger(T) = \hat{\phi}_3^\dagger(0)$ and vice versa, it preserves the cyclicity in spirit. Importantly, the $\hat{\phi}_k^\dagger(t)$ do no propagate according to any physical equation of motion.

Let us look in detail at the single-photon holonomy depicted in Fig. 3b, using the basis $\{\hat{\phi}_1^\dagger, \hat{\phi}_4^\dagger\}$. The first matrix element of the gauge field $\hat{A}$ for this subspace reads (all time-dependencies are omitted, parentheses are used to indicate the action of the time derivative $\mathrm{d}t$ and e is Euler's number):

$$
\begin{aligned}
A_{1,1} &= \mathrm{i}\,\langle 0|\hat{\phi}_1\,\mathrm{d}t\,\hat{\phi}_1^\dagger|0\rangle \\
&= \mathrm{i}\,\langle 0|\mathrm{e}^{\mathrm{i}\delta/2}\,\hat{a}_1\,\mathrm{d}t\big(\mathrm{e}^{-\mathrm{i}\delta/2}\,\hat{a}_1^\dagger\big)|0\rangle \\
&= \mathrm{i}\,\langle 0|\mathrm{e}^{\mathrm{i}\delta/2}\,\hat{a}_1\,(-\mathrm{i}\,\Omega/2)\,\mathrm{e}^{-\mathrm{i}\delta/2}\,\hat{a}_1^\dagger|0\rangle + \mathrm{i}\,\langle 0|\mathrm{e}^{\mathrm{i}\delta/2}\,\hat{a}_1\,\mathrm{e}^{-\mathrm{i}\delta/2}\,\mathrm{d}t\hat{a}_1^\dagger\big)|0\rangle \\
&= \frac{\Omega}{2}
\end{aligned}
$$

In the last step, we used the Heisenberg equation of motion $\mathrm{i}\,\mathrm{d}t\hat{a}_1^\dagger = \hat{a}_1^\dagger\hat{H} - \hat{H}\hat{a}_1^\dagger$, $\hat{H}|0\rangle = 0$ and the fact, that the dynamical phase vanishes, i.e. $\langle 0|\,\hat{a}_1\,\hat{H}\hat{a}_1^\dagger|0\rangle = 0$. The other elements of $\hat{A}$ can be calculated similarly, leading to

$$
\hat{A}_\phi = \begin{pmatrix} \frac{\Omega}{2} & 0 \\ 0 & \frac{\Omega}{2} \end{pmatrix}
$$

in the basis $\{\hat{\phi}_1^\dagger, \hat{\phi}_4^\dagger\}$ and the holonomy

$$
\hat{\mathcal{U}}_a = \begin{pmatrix} 0 & \mathrm{i} \\ \mathrm{i} & 0 \end{pmatrix}
$$

in the basis $\{\hat{a}_1^\dagger, \hat{a}_4^\dagger\}$.

Regarding the two-photon holonomies depicted in Fig. 3 and Fig. 4, the calculation of the gauge fields and holonomies follows a similar vein. However, normalization factors that stem from terms like $\hat{a}_1^\dagger\hat{a}_1^\dagger|0000\rangle = \sqrt{2}|2000\rangle$ (in a photon-number basis) have to be considered. Let us again consider the calculation of one element in detail:

$$
\begin{aligned}
A_{11,11} &= \frac{\mathrm{i}}{2}\,\langle 0|\hat{\phi}_1\hat{\phi}_1\,\mathrm{d}t\,\hat{\phi}_1^\dagger\hat{\phi}_1^\dagger|0\rangle \\
&= \frac{\mathrm{i}}{2}\,\langle 0|\mathrm{e}^{\mathrm{i}\delta}\,\hat{a}_1\hat{a}_1\,\mathrm{d}t\big(\mathrm{e}^{-\mathrm{i}\delta}\,\hat{a}_1^\dagger\hat{a}_1^\dagger\big)|0\rangle \\
&= \frac{\mathrm{i}}{2}\,\langle 0|\mathrm{e}^{\mathrm{i}\delta}\,\hat{a}_1\hat{a}_1\,(-\mathrm{i}\Omega)\,\mathrm{e}^{-\mathrm{i}\delta}\,\hat{a}_1^\dagger\hat{a}_1^\dagger|0\rangle + \frac{\mathrm{i}}{2}\,\langle 0|\mathrm{e}^{\mathrm{i}\delta}\,\hat{a}_1\hat{a}_1\,\mathrm{e}^{-\mathrm{i}\delta}\,\mathrm{d}t\big(\,\hat{a}_1^\dagger\hat{a}_1^\dagger\big)|0\rangle \\
&= \Omega
\end{aligned}
$$

With similar calculations, we find the following results:

| Subspace in $\phi$-Basis | Gauge Field $\hat{A}_\phi$ | Subspace in $a$-Basis | Holonomy $\hat{\mathcal{U}}_a$ |
|---|---|---|---|
| $\{\hat{\phi}_1^\dagger\hat{\phi}_1^\dagger, \hat{\phi}_1^\dagger\hat{\phi}_4^\dagger, \hat{\phi}_4^\dagger\hat{\phi}_4^\dagger\}$ | $\begin{pmatrix} \Omega & 0 & 0 \\ 0 & \Omega & 0 \\ 0 & 0 & \Omega \end{pmatrix}$ | $\{\hat{a}_1^\dagger\hat{a}_1^\dagger, \hat{a}_1^\dagger\hat{a}_4^\dagger, \hat{a}_4^\dagger\hat{a}_4^\dagger\}$ | $\begin{pmatrix} 0 & 0 & -1 \\ 0 & -1 & 0 \\ -1 & 0 & 0 \end{pmatrix}$ |
| $\{\hat{\phi}_1^\dagger\hat{\phi}_1^\dagger, \hat{\phi}_4^\dagger\hat{\phi}_4^\dagger\}$ | $\begin{pmatrix} \Omega & 0 \\ 0 & \Omega \end{pmatrix}$ | $\{\hat{a}_1^\dagger\hat{a}_1^\dagger, \hat{a}_4^\dagger\hat{a}_4^\dagger\}$ | $\begin{pmatrix} 0 & -1 \\ -1 & 0 \end{pmatrix}$ |
| $\{\hat{\phi}_2^\dagger\hat{\phi}_2^\dagger, \hat{\phi}_3^\dagger\hat{\phi}_3^\dagger\}$ | $\begin{pmatrix} \Omega & 0 \\ 0 & \Omega \end{pmatrix}$ | $\{\hat{a}_2^\dagger\hat{a}_2^\dagger, \hat{a}_3^\dagger\hat{a}_3^\dagger\}$ | $\begin{pmatrix} 0 & -1 \\ -1 & 0 \end{pmatrix}$ |
| $\{\hat{\phi}_1^\dagger\hat{\varphi}_3^\dagger, \hat{\phi}_2^\dagger\hat{\varphi}_1^\dagger, \hat{\phi}_3^\dagger\hat{\varphi}_4^\dagger, \hat{\phi}_4^\dagger\hat{\varphi}_2^\dagger\}$ | $\begin{pmatrix} \Omega & 0 & 0 & 0 \\ 0 & \Omega & 0 & 0 \\ 0 & 0 & \Omega & 0 \\ 0 & 0 & 0 & \Omega \end{pmatrix}$ | $\{\hat{a}_1^\dagger\hat{b}_3^\dagger, \hat{a}_2^\dagger\hat{b}_1^\dagger, \hat{a}_3^\dagger\hat{b}_4^\dagger, \hat{a}_4^\dagger\hat{b}_2^\dagger\}$ | $\begin{pmatrix} 0 & 0 & 0 & -1 \\ 0 & 0 & -1 & 0 \\ 0 & -1 & 0 & 0 \\ -1 & 0 & 0 & 0 \end{pmatrix}$ |

Here, we used $\hat{\phi}_k^\dagger$ and $\hat{a}_k^\dagger$ for the first type of particle if the particles are distinguishable and $\hat{\varphi}_k^\dagger$ and $\hat{b}_k^\dagger$ for the second type. See supplementary text for more holonomies that arise in this waveguide structure. The waveguide basis $\hat{a}_k^\dagger$ is one specific example of a propagating basis $\hat{\psi}_k^\dagger$, which is convenient to use in this specific system.

## Derivation of the two-particle holonomic condition

The standard commutator relations read $\hat{a}_k\hat{a}_j^\dagger = \delta_{kj} + \hat{a}_j^\dagger\hat{a}_k$ for bosons and $\hat{a}_k\hat{a}_j^\dagger = \delta_{kj} - \hat{a}_j^\dagger\hat{a}_k$ for fermions. Here, $\delta_{kj}$ is the Kronecker-delta with $\delta_{kk} = 1$ and $\delta_{kj} = 0$ if $k \neq j$. Let us use a general commutator relation $\hat{a}_k\hat{a}_j^\dagger = \delta_{kj} \pm \hat{a}_j^\dagger\hat{a}_k$ for any particle type. In the following derivation, the upper sign will always refer to bosons and the lower sign to fermions. In this section, we assume a linear and Hermitian coupled mode Hamiltonian $\hat{H} = \sum_{k,j}\kappa_{jk}\,\hat{a}_k^\dagger\hat{a}_j$. The modes $\hat{\phi}_A$ that form the non-dynamical and (usually) cyclic basis, at any point in time, are a superposition of propagating modes. So let us express them as $\hat{\phi}_A = \sum_k A_k\hat{a}_k$, where $A_k = A_k(t) \in \mathbb{C}$, then:

$$\begin{aligned}
K_{AB,CD} &= \mathcal{N}\langle 0|\hat{\phi}_A\hat{\phi}_B\hat{H}\hat{\phi}_C^\dagger\hat{\phi}_D^\dagger|0\rangle \\
&= \mathcal{N}\sum_{klnmpq} A_nB_m\kappa_{lk}C_p^*D_q^*\langle 0|\hat{a}_n\hat{a}_m\,\hat{a}_k^\dagger\hat{a}_l\,\hat{a}_p^\dagger\hat{a}_q^\dagger|0\rangle \\
&= \mathcal{N}\sum_{klnmpq} A_nB_m\kappa_{lk}C_p^*D_q^*\langle 0|\hat{a}_n\,(\delta_{mk} \pm\, \hat{a}_k^\dagger\hat{a}_m)(\delta_{lp} \pm\, \hat{a}_p^\dagger\hat{a}_l)\,\hat{a}_q^\dagger|0\rangle \\
&= \mathcal{N}\sum_{klnmpq} A_nB_m\kappa_{lk}C_p^*D_q^*\big[\delta_{mk}\delta_{lp}\langle 0|\hat{a}_n\hat{a}_q^\dagger|0\rangle \pm \delta_{mk}\langle 0|\hat{a}_n\hat{a}_p^\dagger\hat{a}_l\hat{a}_q^\dagger|0\rangle \\
&\qquad \pm\, \delta_{lp}\langle 0|\hat{a}_n\hat{a}_k^\dagger\hat{a}_m\hat{a}_q^\dagger|0\rangle + \langle 0|\hat{a}_n\hat{a}_k^\dagger\hat{a}_m\hat{a}_p^\dagger\hat{a}_l\hat{a}_q^\dagger|0\rangle\big]
\end{aligned}$$

Here, $\mathcal{N}$ is a normalization factor that stems from $\hat{a}_k^\dagger\hat{a}_k^\dagger|0\rangle = \sqrt{2}|2\rangle$ for bosons. Let us look at two of these terms more closely:

$$\begin{aligned}
\langle 0|\hat{a}_n\hat{a}_p^\dagger\hat{a}_l\hat{a}_q^\dagger|0\rangle &= \langle 0|(\delta_{np} \pm \hat{a}_p^\dagger\hat{a}_n)\,(\delta_{lq} \pm \hat{a}_q^\dagger\hat{a}_l)|0\rangle \\
&= \delta_{np}\delta_{lq}
\end{aligned}$$

using $\hat{a}_n|0\rangle = 0$.

$$\begin{aligned}\langle 0|\hat{a}_n\hat{a}_k^\dagger\hat{a}_m\hat{a}_p^\dagger\hat{a}_l\hat{a}_q^\dagger|0\rangle &= \langle 0|(\delta_{nk} \pm \hat{a}_k^\dagger\hat{a}_n)\,\hat{a}_m\hat{a}_p^\dagger\,(\delta_{lq} \pm \hat{a}_q^\dagger\hat{a}_l)|0\rangle \\ &= \delta_{nk}\delta_{lq}\langle 0|\hat{a}_m\hat{a}_p^\dagger|0\rangle \\ &= \delta_{nk}\delta_{lq}\delta_{mp}\end{aligned}$$

using $\langle 0|\hat{a}_k^\dagger = 0$ as well and the assumed orthogonality of the basis. With this

$$K_{AB,CD} = \mathcal{N}\sum_{klnmpq} A_n B_m \kappa_{lk} C_p^* D_q^* \left[\delta_{mk}\delta_{lp}\delta_{nq} \pm \delta_{mk}\delta_{np}\delta_{lq} \pm \delta_{lp}\delta_{nk}\delta_{mq} + \delta_{nk}\delta_{lq}\delta_{mp}\right]$$

$$= \mathcal{N}\left[\sum_{klnq} \delta_{nq} A_n B_k \kappa_{lk} C_l^* D_q^* \pm \sum_{klnp} \delta_{np} A_n B_k \kappa_{lk} C_p^* D_l^* \pm \delta_{mq}\sum_{klmq} \delta_{mq} A_k B_m \kappa_{lk} C_l^* D_q^* \right.$$
$$\left. + \sum_{klmp} \delta_{mp} A_k B_m \kappa_{lk} C_p^* D_l^* \right]$$

On the other hand:

$$\begin{aligned}\mathcal{J}_{AC} &= \langle 0|\hat{\phi}_A \hat{H} \hat{\phi}_C^\dagger|0\rangle \\ &= \sum_{klnp} A_n \kappa_{lk} C_p^* \langle 0|\hat{a}_n\, \hat{a}_k^\dagger \hat{a}_l\, \hat{a}_p^\dagger|0\rangle \\ &= \sum_{klnp} A_n \kappa_{lk} C_p^* \langle 0|\left(\delta_{nk} \pm \hat{a}_k^\dagger\hat{a}_n\right)\left(\delta_{lp} \pm \hat{a}_p^\dagger\hat{a}_l\right)|0\rangle \\ &= \sum_{klnp} A_n \kappa_{lk} C_p^*\, \delta_{nk}\delta_{lp} \\ &= \sum_{kl} A_k \kappa_{lk} C_l^*\end{aligned}$$

and due to orthogonality

$$\begin{aligned}\delta_{AC} &= \langle 0|\hat{\phi}_A\hat{\phi}_C^\dagger|0\rangle \\ &= \sum_{np} A_n C_p^* \langle 0|\hat{a}_n\, \hat{a}_p^\dagger|0\rangle \\ &= \sum_{np} A_n C_p^*\, \delta_{np}.\end{aligned}$$

Using these expressions, we obtain

$$\begin{aligned}K_{AB,CD} &= \mathcal{N}\langle 0|\hat{\phi}_A\hat{\phi}_B\hat{H}\hat{\phi}_C^\dagger\hat{\phi}_D^\dagger|0\rangle \\ &= \mathcal{N}[\delta_{AD}\mathcal{J}_{BC} \pm \delta_{AC}\mathcal{J}_{BD} \pm \delta_{BD}\mathcal{J}_{AC} + \delta_{BC}\mathcal{J}_{AD}].\end{aligned}$$

Hence, the two-particle holonomic condition read $K_{AB,CD} = 0$:

**indistinguishable bosons**: $\delta_{AD}\mathcal{J}_{BC} + \delta_{AC}\mathcal{J}_{BD} + \delta_{BD}\mathcal{J}_{AC} + \delta_{BC}\mathcal{J}_{AD} = 0$

**indistinguishable fermions**: $\delta_{AD}\mathcal{J}_{BC} - \delta_{AC}\mathcal{J}_{BD} - \delta_{BD}\mathcal{J}_{AC} + \delta_{BC}\mathcal{J}_{AD} = 0$

Similarly, we can find a holonomic condition for two **distinguishable particles**. For example,

$$K_{AB,CD} = \langle 0|\hat{\phi}_A \hat{\varphi}_B \hat{H} \hat{\varphi}_C^\dagger \hat{\phi}_D^\dagger |0\rangle = \delta_{AD}\mathcal{J}_{BC} + \delta_{BC}\mathcal{J}_{AD} = 0,$$

where $\hat{\phi}_{A,D}$ refers to one particle type and $\hat{\varphi}_{B,C}$ the other and this condition is the same for both bosons and fermions.

A general expression for a two-particle gauge field

In a similar vein, a general expression for a two-particle gauge field $\hat{A}_\phi$ can be found. In the following, we use parenthesis to clearly indicate onto which parts the differential operator $\mathrm{d}t$ acts.

$$\begin{aligned}
A_{AB,CD} &= \mathcal{N}\,\mathrm{i}\,\langle 0|\hat{\phi}_A\hat{\phi}_B\,\mathrm{d}t\big[\hat{\phi}_C^\dagger\hat{\phi}_D^\dagger\big]|0\rangle \\
&= \mathcal{N}\,\mathrm{i}\,\langle 0|\hat{\phi}_A\hat{\phi}_B\,\mathrm{d}t\big[\hat{\phi}_C^\dagger\big]\hat{\phi}_D^\dagger|0\rangle + \mathcal{N}\,\mathrm{i}\,\langle 0|\hat{\phi}_A\hat{\phi}_B\,\hat{\phi}_C^\dagger\mathrm{d}t\big[\hat{\phi}_D^\dagger\big]|0\rangle \\
&= \pm\,\mathcal{N}\,\mathrm{i}\,\langle 0|\hat{\phi}_A\hat{\phi}_B\,\hat{\phi}_D^\dagger\mathrm{d}t\big[\hat{\phi}_C^\dagger\big]|0\rangle + \mathcal{N}\,\mathrm{i}\,\langle 0|\hat{\phi}_A\hat{\phi}_B\,\hat{\phi}_C^\dagger\mathrm{d}t\big[\hat{\phi}_D^\dagger\big]|0\rangle \\
&= \pm\,\mathcal{N}\,\mathrm{i}\,\langle 0|\hat{\phi}_A\big(\delta_{BD} \pm \hat{\phi}_D^\dagger\hat{\phi}_B\big)\mathrm{d}t\big[\hat{\phi}_C^\dagger\big]|0\rangle + \mathcal{N}\,\mathrm{i}\,\langle 0|\hat{\phi}_A\big(\delta_{BC} \pm \hat{\phi}_C^\dagger\hat{\phi}_B\big)\mathrm{d}t\big[\hat{\phi}_D^\dagger\big]|0\rangle \\
&= \pm\,\mathcal{N}\,\mathrm{i}\,\delta_{BD}\langle 0|\hat{\phi}_A\mathrm{d}t\big[\hat{\phi}_C^\dagger\big]|0\rangle + \mathcal{N}\,\mathrm{i}\;\delta_{BC}\langle 0|\hat{\phi}_A\mathrm{d}t\big[\hat{\phi}_D^\dagger\big]|0\rangle + \mathcal{N}\,\mathrm{i}\,\langle 0|\hat{\phi}_A\hat{\phi}_D^\dagger\hat{\phi}_B\mathrm{d}t\big[\hat{\phi}_C^\dagger\big]|0\rangle \\
&\qquad \pm\,\mathcal{N}\,\mathrm{i}\,\langle 0|\hat{\phi}_A\hat{\phi}_C^\dagger\hat{\phi}_B\mathrm{d}t\big[\hat{\phi}_D^\dagger\big]|0\rangle
\end{aligned}$$

using $\hat{\phi}_C^\dagger\hat{\phi}_D^\dagger = \pm\hat{\phi}_D^\dagger\hat{\phi}_C^\dagger$. Let us have a closer look at the term

$$\begin{aligned}
&\langle 0|\hat{\phi}_A\hat{\phi}_D^\dagger\hat{\phi}_B\mathrm{d}t\big[\hat{\phi}_C^\dagger\big]|0\rangle = \langle 0|\big(\delta_{AD} \pm \hat{\phi}_D^\dagger\hat{\phi}_A\big)\hat{\phi}_B\mathrm{d}t\big[\hat{\phi}_C^\dagger\big]|0\rangle \\
&= \delta_{AD}\langle 0|\hat{\phi}_B\mathrm{d}t\big[\hat{\phi}_C^\dagger\big]|0\rangle.
\end{aligned}$$

This leads us to the final expression for the relation between elements of two-particle gauge fields $A_{AB,CD}$ and elements of single-particle gauge fields $A_{AC}$:

$$A_{AB,CD} = \mathcal{N}\big[\pm\,\delta_{BD}A_{AC} + \delta_{BC}A_{AD} + \delta_{AD}A_{BC} \pm \delta_{AC}A_{BD}\big]$$

Holonomic condition for N-bosons

Let us assume a linear and Hermitian Hamiltonian $\hat{H}$ without further assumptions. Due to linearity

$$\langle 0|\hat{\phi}_{k_1} \dots \hat{\phi}_{k_N}\hat{H}\hat{\phi}_{l_M}^\dagger \dots \hat{\phi}_{l_1}^\dagger|0\rangle = 0,$$

if $N \neq M$. For subspaces with constant particle number $N$ (i.e. $N = M$), the holonomic condition for $N$ bosons reads:

$$\begin{aligned}
K^{(N)}_{k_1\dots k_N, l_1\dots l_N} &= \mathcal{N}\langle 0|\hat{\phi}_{k_1} \dots \hat{\phi}_{k_N}\hat{H}\hat{\phi}_{l_1}^\dagger \dots \hat{\phi}_{l_N}^\dagger|0\rangle \\
&= \mathcal{N} \sum_{\mu=1}^{N}\sum_{\nu=1}^{N}\mathcal{J}_{k_\nu l_\mu} \sum_{\substack{\text{all } \pi(\beta) \\ \beta=1\dots N \\ \beta\neq\mu}} \prod_{\substack{\alpha=1 \\ \alpha\neq\nu}}^{N} \delta_{k_\alpha l_{\pi(\beta)}} - (N-1)\,\langle \hat{H}\rangle \sum_{\substack{\text{all } \pi(\beta) \\ \beta=1\dots N}} \prod_{\alpha=1}^{N} \delta_{k_\alpha l_{\pi(\beta)}}
\end{aligned}$$

with $\mathcal{N}$ being a normalization factor stemming from terms like $\hat{a}_k^\dagger \hat{a}_k^\dagger |0\rangle = \sqrt{2}|2\rangle$, $\mathcal{J}_{k_\nu l_\mu} = \left\langle 0 \middle| \hat{\phi}_{k_\nu} \hat{H} \hat{\phi}_{l_\mu}^\dagger \middle| 0 \right\rangle$ and $\langle \hat{H} \rangle = \langle 0|\hat{H}|0\rangle$. The sum over all $\pi(\beta)$ refers to all possible permutations. Let us proof this condition by induction:

<u>$N = 1$</u>

$$K^{(1)}_{k_1,l_1} = \mathcal{J}_{k_1 l_1}$$

is trivial.

<u>$N = 2$</u>

Let us assume, that there is a basis of propagating modes $\hat{a}_k$ (with no explicit time-dependency) such that the modes fulfill the Heisenberg equation of motion $\mathrm{i}\mathrm{d}_t \hat{a}_k = \hat{a}_k \hat{H} - \hat{H}\hat{a}_k$. For readability, we will omit the hats on operators $\hat{a}_k$ and $\hat{H}$ and indicate a time-derivative with a dot $\dot{a}_k = \mathrm{d}_t \hat{a}_k$.

$$\begin{aligned}
K^{(2)}_{k_1 k_2, l_1 l_2} &= \mathcal{N} \langle 0 | \hat{\phi}_{k_1} \hat{\phi}_{k_2} \hat{H} \hat{\phi}^\dagger_{l_1} \hat{\phi}^\dagger_{l_2} | 0 \rangle \\
&= \mathcal{N} \sum_{k_1, k_2, l_1, l_2} A_{k_1} A_{k_2} A^*_{l_1} A^*_{l_2} \langle 0 | a_{k_1} a_{k_2} \, H \, a^\dagger_{l_1} a^\dagger_{l_2} | 0 \rangle \\
&= \mathcal{N} \sum_{k_1, k_2, l_1, l_2} A_{k_1} A_{k_2} A^*_{l_1} A^*_{l_2} \left( \mathrm{i} \langle 0 | a_{k_1} \dot{a}_{k_2} \, a^\dagger_{l_1} a^\dagger_{l_2} | 0 \rangle + \langle 0 | a_{k_1} H \;\, a_{k_2} \, a^\dagger_{l_1} a^\dagger_{l_2} | 0 \rangle \right)
\end{aligned}$$

Due to linearity $\dot{a}_k a_l = a_l \dot{a}_k$.

$$\begin{aligned}
K^{(2)}_{k_1 k_2, l_1 l_2} &= \mathcal{N} \sum_{k_1, k_2, l_1, l_2} A_{k_1} A_{k_2} A^*_{l_1} A^*_{l_2} \\
&\left( \mathrm{i} \langle 0 | \dot{a}_{k_2} a_{k_1} \, a^\dagger_{l_1} a^\dagger_{l_2} | 0 \rangle + \delta_{k_2 l_1} \langle 0 | a_{k_1} H \;\, a^\dagger_{l_2} | 0 \rangle + \langle 0 | a_{k_1} H \;\, a^\dagger_{l_1} a_{k_2} \, a^\dagger_{l_2} | 0 \rangle \right) \\
&= \mathcal{N} \sum_{k_1, k_2, l_1, l_2} A_{k_1} A_{k_2} A^*_{l_1} A^*_{l_2} \\
&\left( \mathrm{i} \, \delta_{k_1 l_1} \langle 0 | \dot{a}_{k_2} a^\dagger_{l_2} | 0 \rangle + \mathrm{i} \, \delta_{k_1 l_2} \langle 0 | \dot{a}_{k_2} a^\dagger_{l_1} | 0 \rangle + \delta_{k_2 l_1} \langle 0 | a_{k_1} H \;\, a^\dagger_{l_2} | 0 \rangle + \delta_{k_2 l_2} \langle 0 | a_{k_1} H \;\, a^\dagger_{l_1} | 0 \rangle \right)
\end{aligned}$$

Using the Heisenberg equation of motion again, we see that

$$\begin{aligned}
\mathrm{i} \, \langle 0 | \dot{a}_{k_2} a^\dagger_{l_2} | 0 \rangle &= \langle 0 | a_{k_2} H \;\, a^\dagger_{l_2} | 0 \rangle - \langle 0 | \, H \, a_{k_2} \, a^\dagger_{l_2} | 0 \rangle \\
&= \langle 0 | a_{k_2} H \;\, a^\dagger_{l_2} | 0 \rangle - \delta_{k_2 l_2} \langle 0 | \hat{H} | 0 \rangle.
\end{aligned}$$

Hence,

$$\begin{aligned}
K^{(2)}_{k_1 k_2, l_1 l_2} = \mathcal{N} \big( & \delta_{k_1 l_1} \mathcal{J}_{k_2 l_2} + \delta_{k_1 l_2} \mathcal{J}_{k_2 l_1} + \delta_{k_2 l_1} \mathcal{J}_{k_1 l_2} + \delta_{k_2 l_2} \mathcal{J}_{k_1 l_1} - \delta_{k_1 l_1} \delta_{k_2 l_2} \langle \hat{H} \rangle \\
& - \delta_{k_1 l_2} \delta_{k_2 l_1} \langle \hat{H} \rangle \big).
\end{aligned}$$

This result is consistent with the one from a previous section, as for a coupled mode Hamiltonian $\hat{H} = \sum_{k,j} \kappa_{jk} \, \hat{a}_k^\dagger \hat{a}_j$ the expectation value of the vacuum energy vanishes, $\langle 0 | \hat{H} | 0 \rangle = 0$.

$\underline{N-1\rightarrow N}$

$$
\begin{aligned}
K^{(N)}_{k_1\ldots k_N,l_1\ldots l_N} &= \mathcal{N}\langle 0|\hat{\phi}_{k_1}\ldots\hat{\phi}_{k_N}\hat{H}\hat{\phi}^\dagger_{l_N}\ldots\hat{\phi}^\dagger_{l_1}|0\rangle \\
&= \mathcal{N}\sum_{k_1\ldots k_N,l_1\ldots l_N} A_{k_1}\ldots A_{k_N}A^*_{l_1}\ldots A^*_{l_N}\langle 0|a_{k_1}\ldots a_{k_N}\, H\, a^\dagger_{l_N}\ldots a^\dagger_{l_1}|0\rangle \\
&= \mathcal{N}\sum_{k_1\ldots k_N,l_1\ldots l_N} A_{k_1}\ldots A_{k_N}A^*_{l_1}\ldots A^*_{l_N}\big[\mathrm{i}\langle 0|a_{k_1}\ldots a_{k_{N-1}}\dot{a}_{k_N}\, a^\dagger_{l_N}\ldots a^\dagger_{l_1}|0\rangle \\
&\qquad + \langle 0|a_{k_1}\ldots a_{k_{N-1}}H\;\; a_{k_N}\, a^\dagger_{l_N}\ldots a^\dagger_{l_1}|0\rangle\big] \\
&= \mathcal{N}\sum_{k_1\ldots k_N,l_1\ldots l_N} A_{k_1}\ldots A_{k_N}A^*_{l_1}\ldots A^*_{l_N}\big[\mathrm{i}\langle 0|\dot{a}_{k_N}a_{k_1}\ldots a_{k_{N-1}}\, a^\dagger_{l_N}\ldots a^\dagger_{l_1}|0\rangle \\
&\qquad + \delta_{k_N l_N}\langle 0|a_{k_1}\ldots a_{k_{N-1}}H\;\; a^\dagger_{l_{N-1}}\ldots a^\dagger_{l_1}|0\rangle \\
&\qquad + \langle 0|a_{k_1}\ldots a_{k_{N-1}}H\;\; a^\dagger_{l_N}a_{k_N}\, a^\dagger_{l_{N-1}}\ldots a^\dagger_{l_1}|0\rangle\big] \\
&= \mathcal{N}\sum_{k_1\ldots k_N,l_1\ldots l_N} A_{k_1}\ldots A_{k_N}A^*_{l_1}\ldots A^*_{l_N}\big[\mathrm{i}\langle 0|\dot{a}_{k_N}a_{k_1}\ldots a_{k_{N-1}}\, a^\dagger_{l_N}\ldots a^\dagger_{l_1}|0\rangle \\
&\qquad + \delta_{k_N l_N}\langle 0|a_{k_1}\ldots a_{k_{N-1}}H\;\; a^\dagger_{l_{N-1}}\ldots a^\dagger_{l_1}|0\rangle \\
&\qquad + \delta_{k_N l_{N-1}}\langle 0|a_{k_1}\ldots a_{k_{N-1}}H\;\; a^\dagger_{l_N}a^\dagger_{l_{N-2}}\ldots a^\dagger_{l_1}|0\rangle + \cdots \\
&\qquad + \delta_{k_N l_1}\langle 0|a_{k_1}\ldots a_{k_{N-1}}H\;\; a^\dagger_{l_N}\ldots a^\dagger_{l_2}|0\rangle\big] \\
&= \mathcal{N}\Bigg[\delta_{k_N l_N}\, K^{(N-1)}_{k_1\ldots k_{N-1},l_1\ldots l_{N-1}} + \delta_{k_N l_{N-1}}\, K^{(N-1)}_{k_1\ldots k_{N-1},l_1\ldots l_{N-2}l_N} + \cdots + \delta_{k_N l_1}\, K^{(N-1)}_{k_1\ldots k_{N-1},l_2\ldots l_N} \\
&\qquad + \sum_{k_1\ldots k_N,l_1\ldots l_N} A_{k_1}\ldots A_{k_N}A^*_{l_1}\ldots A^*_{l_N}\;\; \mathrm{i}\langle 0|\dot{a}_{k_N}a_{k_1}\ldots a_{k_{N-1}}\, a^\dagger_{l_N}\ldots a^\dagger_{l_1}|0\rangle\Bigg]
\end{aligned}
$$

Let us take a closer look at that last term specifically.

$$
\begin{aligned}
&\sum_{k_1\ldots k_N,l_1\ldots l_N} A_{k_1}\ldots A_{k_N}A^*_{l_1}\ldots A^*_{l_N}\;\; \mathrm{i}\langle 0|\dot{a}_{k_N}a_{k_1}\ldots a_{k_{N-1}}\, a^\dagger_{l_N}\ldots a^\dagger_{l_1}|0\rangle \\
&= \sum_{k_1\ldots k_N,l_1\ldots l_N} A_{k_1}\ldots A_{k_N}A^*_{l_1}\ldots A^*_{l_N}\;\; \mathrm{i}\,\big[\delta_{k_{N-1}l_N}\langle 0|\dot{a}_{k_N}a_{k_1}\ldots a_{k_{N-2}}\, a^\dagger_{l_{N-1}}\ldots a^\dagger_{l_1}|0\rangle \\
&\qquad + \langle 0|\dot{a}_{k_N}a_{k_1}\ldots a_{k_{N-2}}\, a^\dagger_{l_N}a_{k_{N-1}}a^\dagger_{l_{N-1}}\ldots a^\dagger_{l_1}|0\rangle\big] \\
&= \sum_{k_1\ldots k_N,l_1\ldots l_N} A_{k_1}\ldots A_{k_N}A^*_{l_1}\ldots A^*_{l_N}\;\; \mathrm{i}\,\big[\delta_{k_{N-1}l_N}\langle 0|\dot{a}_{k_N}a_{k_1}\ldots a_{k_{N-2}}\, a^\dagger_{l_{N-1}}\ldots a^\dagger_{l_1}|0\rangle \\
&\qquad + \delta_{k_{N-1}l_{N-1}}\langle 0|\dot{a}_{k_N}a_{k_1}\ldots a_{k_{N-2}}\, a^\dagger_{l_N}a^\dagger_{l_{N-1}}a_{k_{N-1}}a^\dagger_{l_{N-2}}\ldots a^\dagger_{l_1}|0\rangle + \cdots \\
&\qquad + \delta_{k_{N-1}l_1}\langle 0|\dot{a}_{k_N}a_{k_1}\ldots a_{k_{N-2}}\, a^\dagger_{l_N}\ldots a^\dagger_{l_2}|0\rangle\big] \\
&= \sum_{k_1\ldots k_N,l_1\ldots l_N} A_{k_1}\ldots A_{k_N}A^*_{l_1}\ldots A^*_{l_N}\;\; \mathrm{i}\,\big[\delta_{k_{N-1}l_N}\delta_{k_{N-2}l_{N-1}}\langle 0|\dot{a}_{k_N}a_{k_1}\ldots a_{k_{N-3}}\, a^\dagger_{l_{N-2}}\ldots a^\dagger_{l_1}|0\rangle \\
&\qquad + \delta_{k_{N-1}l_N}\langle 0|\dot{a}_{k_N}a_{k_1}\ldots a_{k_{N-3}}\, a^\dagger_{l_{N-1}}a_{k_{N-2}}a^\dagger_{l_{N-2}}\ldots a^\dagger_{l_1}|0\rangle + \cdots \\
&\qquad + \delta_{k_{N-1}l_1}\delta_{k_{N-2}l_N}\langle 0|\dot{a}_{k_N}a_{k_1}\ldots a_{k_{N-3}}\, a^\dagger_{l_{N-1}}\ldots a^\dagger_{l_2}|0\rangle \\
&\qquad + \delta_{k_{N-1}l_1}\langle 0|\dot{a}_{k_N}a_{k_1}\ldots a_{k_{N-3}}\, a^\dagger_{l_N}a_{k_{N-2}}a^\dagger_{l_N}\ldots a^\dagger_{l_2}|0\rangle\big] \\
&= \cdots
\end{aligned}
$$

$$= \sum_{k_1\dots k_N, l_1 \dots l_N} A_{k_1} \dots A_{k_N} A^*_{l_1} \dots A^*_{l_N} \; \mathrm{i} \left[ \langle 0 | \dot{a}_{k_N} a^\dagger_{l_1} | 0 \rangle \sum_{\substack{\text{all } \pi(\beta) \\ \beta = 2\dots N}} \delta_{k_1 l_{\pi(\beta)}} \delta_{k_2 l_{\pi(\beta)}} \dots \delta_{k_{N-1} l_{\pi(\beta)}} \right.$$

$$+ \langle 0 | \dot{a}_{k_N} a^\dagger_{l_2} | 0 \rangle \sum_{\substack{\text{all } \pi(\beta) \\ \beta = 1,3\dots N}} \delta_{k_1 l_{\pi(\beta)}} \delta_{k_2 l_{\pi(\beta)}} \dots \delta_{k_{N-1} l_{\pi(\beta)}} + \cdots$$

$$\left. + \langle 0 | \dot{a}_{k_N} a^\dagger_{l_N} | 0 \rangle \sum_{\substack{\text{all } \pi(\beta) \\ \beta = 1\dots N-1}} \delta_{k_1 l_{\pi(\beta)}} \delta_{k_2 l_{\pi(\beta)}} \dots \delta_{k_{N-1} l_{\pi(\beta)}} \right]$$

Using again $\mathrm{i} \langle 0 | \dot{a}_{k_N} a^\dagger_{l_1} | 0 \rangle = \langle 0 | a_{k_N} H \; a^\dagger_{l_1} | 0 \rangle - \delta_{k_N l_1} \langle 0 | \hat{H} | 0 \rangle$.

$$= \left( \mathcal{J}_{k_N l_1} - \delta_{k_N l_1} \langle H \rangle \right) \sum_{\substack{\text{all } \pi(\beta) \\ \beta = 2\dots N}} \prod_{\alpha=1}^{N-1} \delta_{k_\alpha l_{\pi(\beta)}} + \left( \mathcal{J}_{k_N l_2} - \delta_{k_N l_2} \langle H \rangle \right) \sum_{\substack{\text{all } \pi(\beta) \\ \beta = 1,3\dots N}} \prod_{\alpha=1}^{N-1} \delta_{k_\alpha l_{\pi(\beta)}} + \cdots$$

$$+ \left( \mathcal{J}_{k_N l_N} - \delta_{k_N l_N} \langle H \rangle \right) \sum_{\substack{\text{all } \pi(\beta) \\ \beta = 1\dots N-1}} \prod_{\alpha=1}^{N-1} \delta_{k_\alpha l_{\pi(\beta)}}$$

$$= \sum_{\mu=1}^{N} \left( \mathcal{J}_{k_N l_\mu} - \delta_{k_N l_\mu} \langle H \rangle \right) \sum_{\substack{\text{all } \pi(\beta) \\ \beta = 1\dots N \\ \beta \neq \mu}} \prod_{\alpha=1}^{N-1} \delta_{k_\alpha l_{\pi(\beta)}}$$

Hence

$$K^{(N)}_{k_1\dots k_N, l_1 \dots l_N} = \mathcal{N} \left[ \delta_{k_N l_N} \, K^{(N-1)}_{k_1 \dots k_{N-1}, l_1 \dots l_{N-1}} + \cdots + \delta_{k_N l_1} \, K^{(N-1)}_{k_1 \dots k_{N-1}, l_2 \dots l_N} \right.$$

$$\left. + \sum_{\mu=1}^{N} \left( \mathcal{J}_{k_N l_\mu} - \delta_{k_N l_\mu} \langle H \rangle \right) \sum_{\substack{\text{all } \pi(\beta) \\ \beta = 1\dots N \\ \beta \neq \mu}} \prod_{\alpha=1}^{N-1} \delta_{k_\alpha l_{\pi(\beta)}} \right]$$

$$
\begin{aligned}
&= \mathcal{N} \Bigg[ \delta_{k_N l_N} \sum_{\mu=1}^{N-1} \sum_{\nu=1}^{N-1} \mathcal{J}_{k_\nu l_\mu} \sum_{\substack{\text{all } \pi(\beta) \\ \beta=1\ldots N-1 \\ \beta\neq\mu}} \prod_{\substack{\alpha=1 \\ \alpha\neq\nu}}^{N-1} \delta_{k_\alpha l_{\pi(\beta)}} - (N-2)\,\langle \widehat{H} \rangle \sum_{\substack{\text{all } \pi(\beta) \\ \beta=1\ldots N-1}} \prod_{\alpha=1}^{N-1} \delta_{k_\alpha l_{\pi(\beta)}} + \cdots \\
&\qquad + \delta_{k_N l_1} \sum_{\mu=2}^{N} \sum_{\nu=1}^{N-1} \mathcal{J}_{k_\nu l_\mu} \sum_{\substack{\text{all } \pi(\beta) \\ \beta=2\ldots N \\ \beta\neq\mu}} \prod_{\substack{\alpha=1 \\ \alpha\neq\nu}}^{N-1} \delta_{k_\alpha l_{\pi(\beta)}} - (N-2)\,\langle \widehat{H} \rangle \sum_{\substack{\text{all } \pi(\beta) \\ \beta=2\ldots N}} \prod_{\alpha=1}^{N-1} \delta_{k_\alpha l_{\pi(\beta)}} \\
&\qquad + \sum_{\mu=1}^{N} \left( \mathcal{J}_{k_N l_\mu} - \delta_{k_N l_\mu} \langle H \rangle \right) \sum_{\substack{\text{all } \pi(\beta) \\ \beta=1\ldots N \\ \beta\neq\mu}} \prod_{\alpha=1}^{N-1} \delta_{k_\alpha l_{\pi(\beta)}} \Bigg] \\
&= \mathcal{N} \sum_{\mu=1}^{N} \sum_{\nu=1}^{N} \mathcal{J}_{k_\nu l_\mu} \sum_{\substack{\text{all } \pi(\beta) \\ \beta=1\ldots N \\ \beta\neq\mu}} \prod_{\substack{\alpha=1 \\ \alpha\neq\nu}}^{N} \delta_{k_\alpha l_{\pi(\beta)}} - (N-1)\,\langle \widehat{H} \rangle \sum_{\substack{\text{all } \pi(\beta) \\ \beta=1\ldots N}} \prod_{\alpha=1}^{N} \delta_{k_\alpha l_{\pi(\beta)}}
\end{aligned}
$$

□

This proves the general holonomic condition for N bosons in a linear and Hermitian system.

Experimental configuration

To experimentally show the effects of two-particle holonomies, a waveguide system that gives rise to such holonomies is examined with two indistinguishable photons as well as single photons. To create the waveguides ultrashort laser pulses from a frequency-doubled fiber amplifier system (Coherent Monaco, wavelength 517 nm, repetition rate 333 kHz, pulse duration 270 fs) are focused into the bulk of a fused silica chip (Corning 7980, dimensions 1 mm × 20 mm × 100 mm, bulk refractive index $n_0 \approx 1.453$ at 815 nm). The nonlinear absorption and subsequent rapid quenching of the intermittent plasma back to room temperature induces a localized and permanent refractive index change in the focal volume ($\Delta n_0 \approx 2 \times 10^{-3}$). A precision translation system (Aerotech ALS130, operated at: writing speed 100 mm min$^{-1}$) defines the desired waveguide trajectory, resulting in high-quality single-mode waveguides.

The four-waveguide structure should abide the Hamiltonian $\widehat{H} = \kappa_\mathrm{P} \hat{a}_1^\dagger \hat{a}_2 + \kappa_\mathrm{C} \hat{a}_2^\dagger \hat{a}_3 + \kappa_\mathrm{P} \hat{a}_3^\dagger \hat{a}_4 + \mathrm{h.c.}$ with central coupling $\kappa_\mathrm{C} = \Omega(t)$ and peripheral coupling $\kappa_\mathrm{P} = \frac{\sqrt{3}}{2}\Omega(t)$. Using a coupling scan, couplings can be directly related to real space distances between neighboring waveguides. We choose an envelope function with respect to distances

$$\Delta_\Omega(z) = \begin{cases} c_1 \cos\left(\frac{\pi z}{30\text{ mm}}\right) + c_2 & 0 \leq z \leq 30\text{ mm} \\ c_2 - c_1 & 30\text{ mm} < z < L - 30\text{ mm} \\ c_1 \cos\left(\frac{\pi (z - 30\text{ mm})}{30\text{ mm}}\right) + c_2 & L - 30\text{ mm} \leq z \leq L \end{cases}$$

with $c_1 \approx 0.03$ mm and $c_2 \approx 0.05$ mm. Seven instances of this structure are realized that differ in total length $L$. Specifically, the length of the straight section in the middle is altered. With this construction at the ideal length $L_{\text{id}} \approx 84.9$ mm the condition $\int_0^T \Omega(t)\mathrm{d}t = \pi$ is fulfilled and the structure behaves perfectly cyclically. (The propagation length $L$ and time $T$ are directly related due to paraxiality.)

To allow for a smooth incoupling of photons into the waveguides using a standard fiber array, a minor correction to the trajectories of the two outer waveguides is necessary. The fiber array consists of several equally spaced fibers with a distance between neighboring cores of 82 µm. With this choice of geometry $\Delta_\Omega(z)$ waveguides two and three have a distance of 82 µm at the front and end facet. However, the distance between the outer most waveguides and their respective neighbors is slightly larger. Therefore, their trajectories deviate from $\Delta_\Omega(z)$ for the first and last 7.5 mm (see supplementary text). However, as the distance at the point of deviation is approximately 74 µm the changes in coupling are neglectable. At such large distances, couplings are in good approximation zero, i.e. so small that they are not measurable.

Pairs of indistinguishable photons are created through type-I spontaneous parametric down-conversion (*6*) in a bismuth borate crystal ($BiB_3O_6$) pumped with a continuous-wave laser (wavelength 407 nm, power 100 mW). The photons are collected by polarization-maintaining fibers and launched into the waveguide structure using an array of polarization-maintaining fibers. At the output, they are collected by an array of multi-mode fibers that route them to avalanche photodetectors (APD) (Excelitas, detection efficiency of >50%, dark counts of <50 s−1, dead time of 20 ns). Using a correlation card (Becker & Hickl, time bin 164 ps), a coincidence is recorded if both photons are detected withing 5 ns of one another.

Different input and output configurations are used for specific parts of the experiment. For single-photon measurements, one photon is used as a herald and transported directly through a fiber to an APD while the signal photon is launched into one waveguide. At the output, each of the four waveguides are connected to an APD and coincidences between the APD that detects the herald and any of the four APDs that detect the signal are counted. This measurement is repeated for each of the four waveguides in each of the seven structures. The coincidence probabilities for two distinguishable photons are retrieved from these heralded single-photon measurements, effectively distinguishing the two photons with a time-delay on the order of minutes.

To differentiate all possible states involving two indistinguishable photons in the measurement, a fiber-integrated multi-mode 50:50 beam splitter is connected to each of the four waveguide outputs, necessitating a total of eight APDs. If the photons form a bunched state, i.e. if both photons leave the waveguide structure in the same waveguide, there is a 50% chance (for ideal beam splitters), that both APDs that are connected to the same fiber beam splitter detect one photon (see supplementary text for calibration measurements of the beam splitters). Effectively, this setup realizes a photon-number resolving measurement for two photons.

On the input side, two indistinguishable photons can be launched simultaneously into two different waveguides, realizing six of the ten two-photon input states. (See supplementary text for calibration measurements that realize simultaneity.) To realize the remaining four input states, where both photons occupy the same mode, the Hong-Ou-Mandel effect (*7*) is exploited. A fiber-integrated single-mode 50:50 beam splitter is placed between the SPDC-source and the waveguide chip. If both photons enter the beam splitter simultaneously they form a bunched state at the output, that is a superposition between both photons in one output and both photons in the other output. One output is connected to the waveguide structure, while the other output is dumped. The 50% losses with this method are compensated with longer measurements. For each of the seven waveguide structures, each of the ten input states is measured in an individual experiment.

Data evaluation and theoretical model

For this specific system, the success probability (as depicted in Fig. 3 and Fig. 4) amounts to a fidelity in the sense of a Bhattacharyya distance (*8*). Generally, the fidelity for a specific input state is defined as $F = \left|\langle\psi_{\text{out}}^{\text{theo}}|\psi_{\text{out}}^{\text{exp}}\rangle\right|^2$, where $\left|\psi_{\text{out}}^{\text{theo/exp}}\right\rangle$ denote the theoretical and experimental output states, respectively. For a single photon, since the states $\hat{a}_k^\dagger|0\rangle$ form an orthonormal basis, the output states can be expanded as $\left|\psi_{\text{out}}^{\text{theo/exp}}\right\rangle = \sum_j \sqrt{p_{kj}^{\text{theo/exp}}}\hat{a}_j^\dagger|0\rangle$, where $p_{kj}^{\text{theo/exp}}$ denotes the probability of the state $\hat{a}_j^\dagger|0\rangle$ being detected after $\hat{a}_k^\dagger|0\rangle$ was launched into the system. Hence, the fidelity for the input state $\hat{a}_k^\dagger|0\rangle$ is $F_k = \left(\sum_j \sqrt{p_{kj}^{\text{theo}} p_{kj}^{\text{exp}}}\right)^2$. A similar expansion is possible for two photons. Since $p_{kj}^{\text{theo}}$ equals always one for success and zero for failure in this specific system, the fidelity is equal to the success probability.

To calculate the theoretical model, depicted in Fig. 3 and Fig. 4, we assumed experimentally obtained values for the couplings $\kappa_{kj}(z)$. Exploiting paraxiality, the time-evolution operator $\hat{U}(L) = \exp(-\mathrm{i}\int_0^L \hat{H}(z)\mathrm{d}z)$ was evaluated explicitly for propagation lengths between $L = 80\text{mm}$ and $L = 100\text{mm}$. The relevant cyclic subspace was enclosed on the input side by selecting only input states that lie inside the subspace. For single-photons this amounts to the absolute square of specific matrix elements of $\hat{U}(L)$. For two-photons an effective two-photon time-evolution operator can be calculated, a 10x10 matrix for indistinguishable photons and a 16x16 matrix for distinguishable photons. The relevant elements of these matrices were selected. Finally, the cyclic subspace is enclosed at the output by renormalizing the detection probabilities, such that the events that are counted as success added by the events that are counted as failure amount to a total detection probability of 1. No knowledge about the (lack of) dynamical phase is needed for calculating the theoretical model. All cyclic subspaces both holonomic and non-holonomic are calculated using the same method (and exact same code).

For the experimental data, the following considerations inform the uncertainties depicted in Fig. 3 and Fig. 4. The uncertainties of the beam splitters at the output side are orders of magnitude smaller than other contributions to the uncertainties and, hence, are neglected. The uncertainty of the beam splitter at the input side (for inputs, where both photons occupy the same mode) per construction does not influence the measurement results. A dark count measurement showed neglectable coincidence counts. As both input fibers and waveguides are single mode, changing the exact incoupling does not influence the measured probabilities and only influences total counts. However, the exact position of the output-fiber array influences the measured probabilities, since the exact part of the detector surface that absorbs the photon depends on the mode that is excited in the multi-mode output fiber. Hence, the two main contributions to the uncertainties are the exact position of the output fiber array and the probabilistic nature of the detection process. For the former, the optimization process of the fiber-array position was repeated several times and its influence to the detection probability analyzed statistically. For the latter a single Poisson standard deviation of the counts in each channel is taken into account, as is a standard method in quantum-optical experiments.

We define a plateau as a continuous section in the propagation length, where the first derivate of the success probability with regard to the propagation length is smaller than 1.5%/mm. The plateau width is the distance between start and end of the plateau. The theoretical plateau width can be extracted directly from the theoretical model. The cutoff of 1.5%/mm is equivalent to a difference of 5% between two experimental data points. Hence, the start points and end points of experimental plateaus are calculated as the first and last data point, where the difference to the next (or previous) data point is smaller than 5%. All plateau widths that are mentioned in the main text are averages over the plateaus of all input states within a subspace. Due to the low sampling rate, the experimental plateau width has a low accuracy and tends to be an underestimation.

## Supplementary Text

### Four-waveguide structure

The seven instances of the four-waveguide structure described in the methods section are depicted in Fig. S1. To allow for a smooth transition to a standard fiber array, a small deviation from the desired waveguide trajectory is necessary.

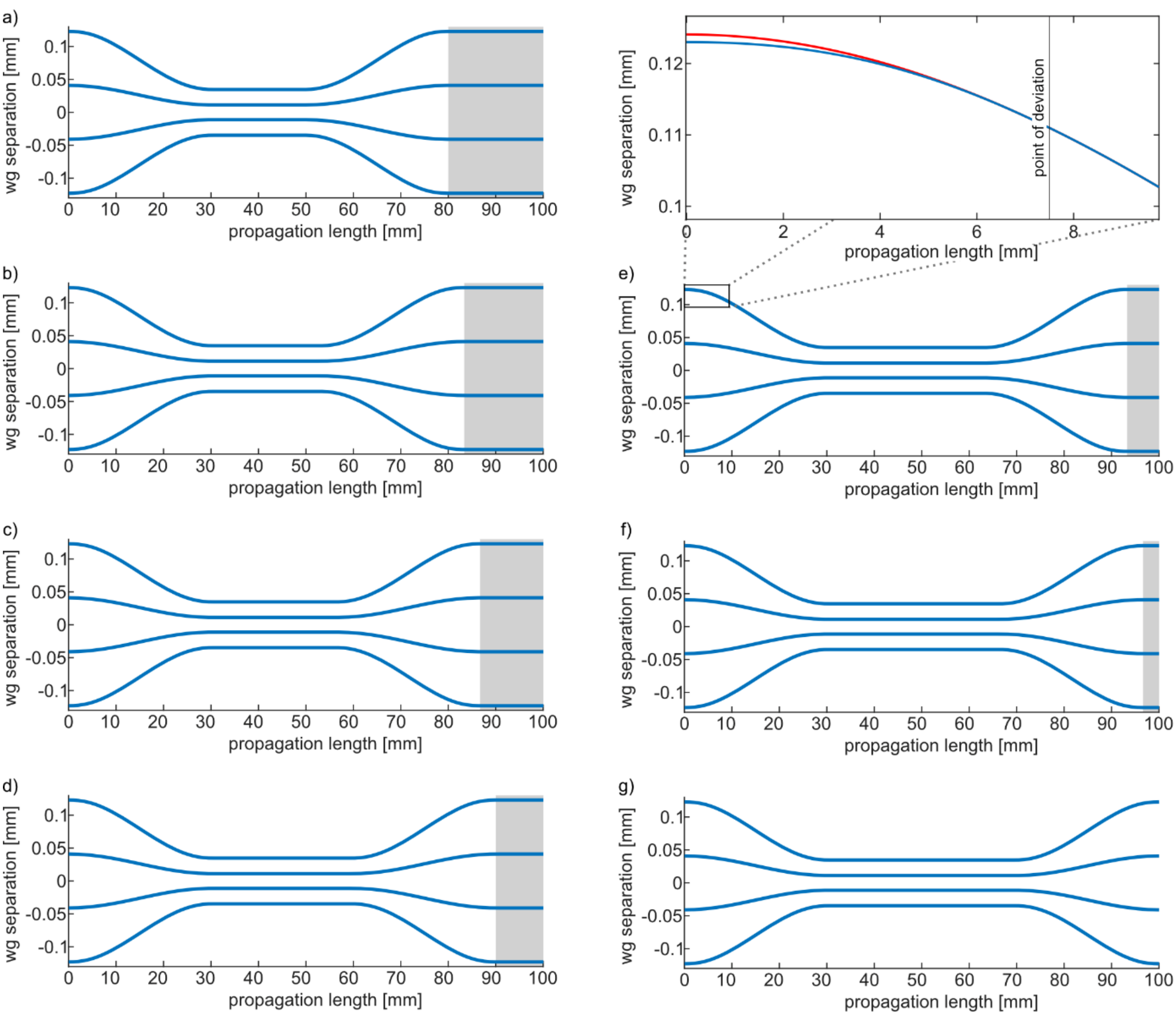

**Figure S1: Geometry of the waveguide structures.** a) – g) show the waveguide separation over propagation length for the seven instances of the four-waveguide structure. The cosine-shaped fanning sections are identical for all seven instances. The straight section in-between varies in length, realizing total propagation lengths between 80mm and 100mm in steps of 1/3cm. As the silica chip has a constant length of 100mm, a straight section at the end of the structure is added (marked in grey). Here, the waveguides are completely decoupled, hence, the straight section does not influence the total evolution. To allow for a smooth transition to a standard fiber array, with a pitch of 82 µm a small correction is necessary. This correction is applied to all of the curved sections of the outer most waveguides both at the front and end facet and depicted in the inset of e). The red curve shows the ideal waveguide trajectory that ensures that the Hamiltonian and the time-evolution operator commute. The blue curve shows the slightly deviating, real waveguide trajectory that ensures a smooth transition to the fiber array. As the coupling between the waveguides at this distance is so small that it cannot be measured, the deviation that this correction introduces is neglectable.

Creation of two-photon input states

Two photons are created through type-I spontaneous parametric down conversion. To prepare a bunched state (i.e. where both photons are launched into the same waveguide), both photons are launched into separate arms of a fiber-integrated beam splitter. Exploiting the Hong-Ou-Mandel (HOM) effect (*7*), a bunched state is created. In one arm, a time delay is introduced using a translation state. By varying the translation, the characteristic HOM-dip is created, see Fig S2. The visibility of 98.6% shows the high indistinguishability of the photons. To launch both photons into the waveguide structure, the translation is set to the lowest point of the dip, and one output of the beam splitter is connected to the waveguide input.

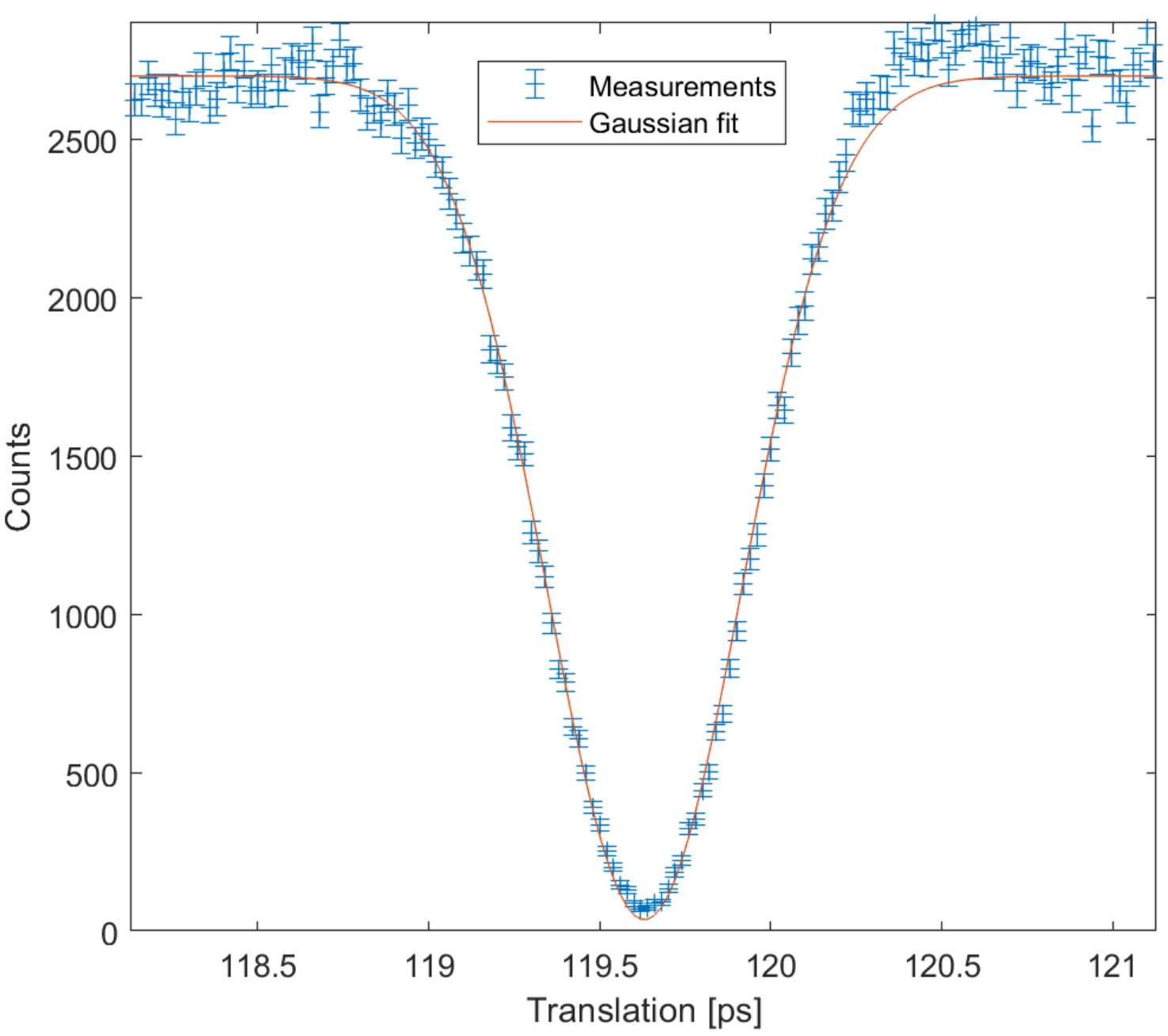

**Figure S2: Hong-Ou-Mandel dip in a fiber-integrated beam splitter.** A time delay is introduced in one arm of the beam splitter. If both photons reach the beam splitter simultaneously, the correlation counts show a minimum. The visibility of 98.6% shows the high indistinguishability of the photons.

For input states, where the photons occupy two different waveguides, it is crucial for them to enter the waveguide chip simultaneously. Inspired by the HOM dip, the two photons were launched into different inputs of structure 7 and a time delay was introduced. At the position of the translation stage where both photons enter the chip simultaneously, either a dip or a peak was observed depending on the input combination (see Fig. S3). The position of this extremum was subsequently used to prepare the input states. This calibration measurement was only performed in structure 7 since the main deviation in path lengths stems from differing lengths of the fibers that lead up to the waveguide structure. Structure 7 was chosen specifically, since it deviates the most from the ideal structure. In the ideal structure, the outputs are expected to be identical for distinguishable and indistinguishable photons. Hence, the dips or peaks should be the most prominent in structure 7.

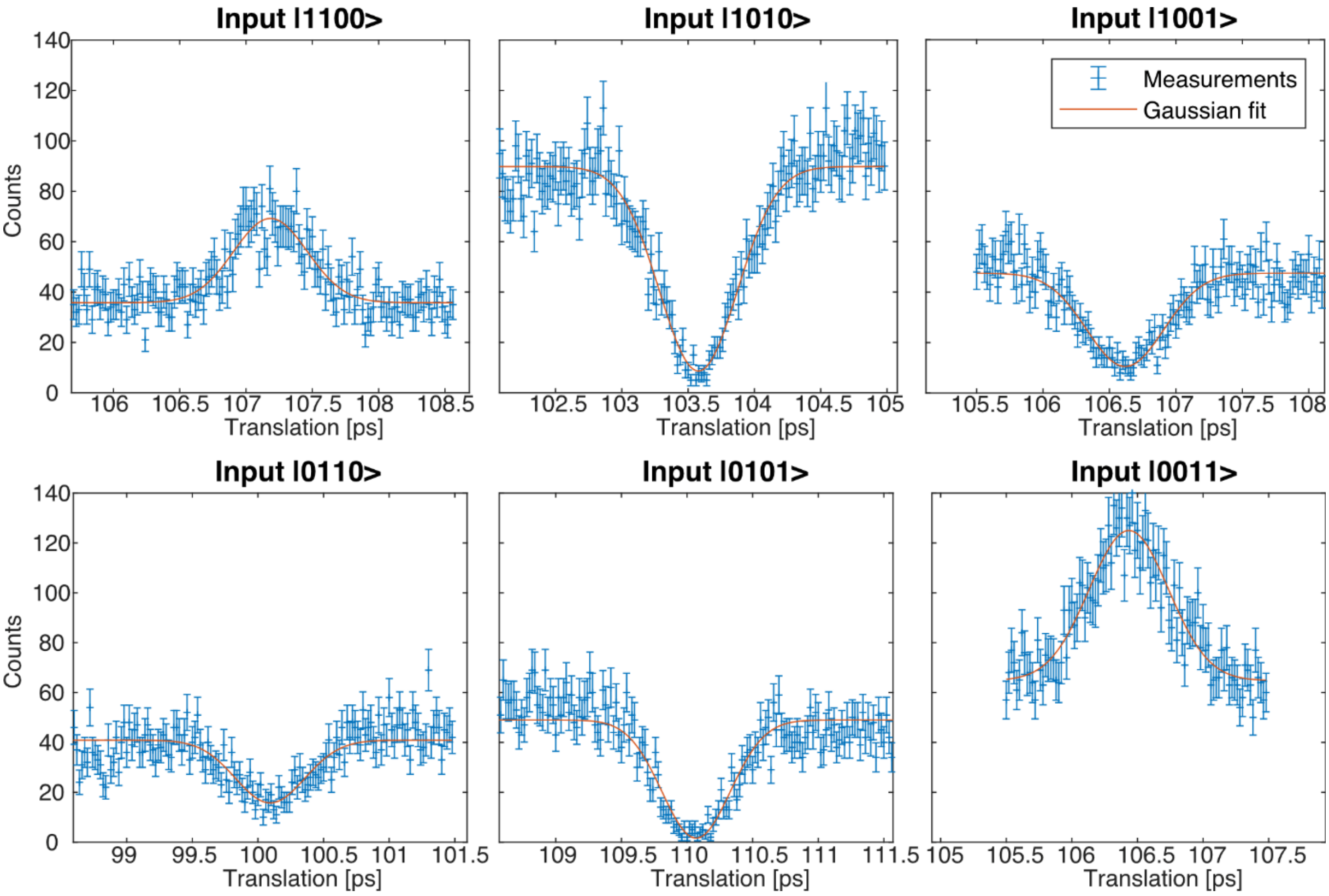

**Figure S3: Calibration measurement to ensure simultaneity.** Two indistinguishable photons are launched into different inputs of structure 7. A time delay between both photons is introduced. At the position of the translation stage where both photons enter the waveguide chip simultaneously a dip or peak is observed in the correlations counts of two arbitrary waveguides at the output. This position can be used to launch the photons into the structure simultaneously.

Beam splitter characterization for photon number-resolving measurement

On the output side, an array of eight avalanche photo diodes detects the photons. Each waveguide is connected to two detectors via a fiber-integrated multi-mode 50:50 beam splitter. This measurement scheme allows for a photon number-resolving measurement. However, the exact splitting ratio of the beam splitters is needed to calculate the results. The splitting ratios are experimentally characterized, by launching single photons directly into the relevant input of the beam splitter and detecting single-photon counts in both outputs, giving the following results.

**Table S1**: Splitting ratio of fiber integrated multi-mode beam-splitters for photon number-resolving measurement

| Output waveguide | Beam splitter output 1 | Beam splitter output 2 |
|---|---|---|
| 1 | $0.5130 \pm 5 \times 10^{-4}$ | $0.4870 \pm 5 \times 10^{-4}$ |
| 2 | $0.5736 \pm 4 \times 10^{-4}$ | $0.4264 \pm 5 \times 10^{-4}$ |
| 3 | $0.4419 \pm 5 \times 10^{-4}$ | $0.5581 \pm 4 \times 10^{-4}$ |
| 4 | $0.4751 \pm 4 \times 10^{-4}$ | $0.5249 \pm 4 \times 10^{-4}$ |

Other holonomies in the four-waveguide structure

The four-waveguide structure, discussed in this work gives rise to a plethora of holonomies due to its high symmetry. Of these holonomies only a small selection is discussed in the main text. Here, we show all 18 non-Abelian holonomies with two indistinguishable photons as well one single-photon holonomy and one holonomy that needs distinguishable photon pairs. In this specific system, every subspace that gives rise to a holonomy with two indistinguishable photons, also gives rise to a holonomy with two distinguishable photons. All of these holonomies are shown in the table below. The first column describes the particle type: single photon (*1 Phot*), indistinguishable (*indist*) or distinguishable (*dist*). The subspace is defined by the input states (expressed as particle number states). The mean plateau width is shown as a general measure for each holonomy. As in the main text, the experimental plateau width is calculated from the measurement data and due to the low sampling rate has a low accuracy and tends to underestimate the width of the plateau. For easy comparison, a mean plateau width is calculated from the theoretical model by restricting the model to only those propagation lengths that have been experimentally evaluated, i.e. from 80 mm to 100 mm. However, some plateaus start before 80 mm. Hence, the second to last column shows the total mean plateau width calculated from the theoretical model, without any restrictions to the propagation length. Interestingly, some very stable holonomies produce mean plateau width of roughly 1/3 of the ideal propagation length, showing ridiculous resilience to changes in propagation length. Furthermore, even the experimental plateau width with its low precision can be used to examine if a subspace is holonomic. For this structure, we can formulate a rule of thumb: a plateau width of less than 3.5 mm indicates a non-holonomic subspace, while a plateau width of more than 6 mm, indicates a holonomy with reasonable confidence. In the range between 3.5 mm and 6 mm the experimental plateau width is not suited to test if a subspace is holonomic.

**Table S2**: Mean plateau width of different holonomies.

| | **Input states** | **Experiment** | **Theory** 80 - 100 mm | **Theory** unrestricted | **In main text** |
|---|---|---|---|---|---|
| **1 Phot** | $\|1000\rangle$, $\|0001\rangle$ | 13 mm | 16.7 mm | 23.7 mm | Fig. 3b |
| **indist** | $\|2000\rangle$, $\|0002\rangle$ | 15 mm | 19.4 mm | 28.9 mm | Fig. 3d |
| **dist** | | 15 mm | 19.4 mm | 28.9 mm | |
| **indist** | $\|0200\rangle$, $\|0020\rangle$ | 10 mm | 7.7 mm | 7.7 mm | Fig. 3g |
| **dist** | | 10 mm | 7.7 mm | 7.7 mm | |
| **indist** | $\|1010\rangle$, $\|0101\rangle$ | 10 mm | 9.5 mm | 9.5 | |
| **dist** | | 10 mm | 9.2 mm | 9.2 | |
| **indist** | $\|1100\rangle$, $\|0011\rangle$ | 10 mm | 12.1 mm | 14.3 mm | |
| **dist** | | 10 mm | 14.3 mm | 18.9 mm | |
| **indist** | $\|2000\rangle$, $\|0002\rangle$, $\|0200\rangle$, | 8 mm | 8.3 mm | 8.3 mm | |
| **dist** | $\|0020\rangle$ | 8 mm | 8.3 mm | 8.3 mm | |
| **indist** | $\|2000\rangle$, $\|0002\rangle$, $\|1010\rangle$, | 6 mm | 9.3 mm | 9.6 mm | |
| **dist** | $\|0101\rangle$ | 7 mm | 9.9 mm | 10.8 mm | |
| **indist** | $\|2000\rangle$, $\|0002\rangle$, $\|0110\rangle$ | 11 mm | 11.4 mm | 12.9 | |
| **dist** | | 11 mm | 12.3 mm | 14.6 | |
| **indist** | $\|2000\rangle$, $\|0002\rangle$, $\|1001\rangle$ | 13 mm | 15.1 mm | 20.2 mm | Fig. 3c |
| **dist** | | 13 mm | 15.9 mm | 21.9 mm | |
| **indist** | $\|0200\rangle$, $\|0020\rangle$, $\|1010\rangle$, | 4 mm | 5.8 mm | 5.8 | |
| **dist** | $\|0101\rangle$ | 4 mm | 6.2 mm | 6.2 | |
| **indist** | $\|0200\rangle$, $\|0020\rangle$, $\|1001\rangle$ | 10 mm | 8.7 mm | 9.2 | |
| **dist** | | 9 mm | 8.8 mm | 9.2 | |
| **indist** | $\|0110\rangle$, $\|1001\rangle$ | 10 mm | 9.1 mm | 9.1 | |
| **dist** | | 8 mm | 9.3 mm | 9.4 | |
| **indist** | $\|0110\rangle$, $\|1100\rangle$, $\|0011\rangle$ | 4 mm | 6.0 mm | 6.0 | |
| **dist** | | 4 mm | 7.6 mm | 7.6 | |
| **indist** | $\|1001\rangle$, $\|1100\rangle$, $\|0011\rangle$ | 8 mm | 10.5 mm | 11.2 | |
| **dist** | | 7 mm | 10.9 mm | 12.0 | |
| **indist** | $\|2000\rangle$, $\|0002\rangle$, $\|0110\rangle$, | 8 mm | 10.5 mm | 11.5 | |
| **dist** | $\|1001\rangle$ | 10 mm | 11.0 mm | 12.5 | |
| **indist** | $\|2000\rangle$, $\|0002\rangle$, $\|0200\rangle$, | 8 mm | 8.8 mm | 9.0 | |
| **dist** | $\|0020\rangle$, $\|1001\rangle$ | 8 mm | 8.8 mm | 9.1 | |
| **indist** | $\|2000\rangle$, $\|0002\rangle$, $\|0200\rangle$, | 4 mm | 6.5 mm | 6.5 mm | |
| **dist** | $\|0020\rangle$, $\|1010\rangle$, $\|0101\rangle$ | 4 mm | 6.9 mm | 6.9 mm | |
| **dist** | $\|A0B0\rangle$, $\|0B0A\rangle$, $\|BA00\rangle$, $\|00AB\rangle$ | 10 mm | 11.5 mm | 13.1 mm | Fig. 4c |

Adiabatic two-photon holonomy

In the main text, the discussion focusses on non-adiabatic holonomies (*2*), as this is a general approach to holonomies. However, the theory of holonomies with more than one particle also applies to adiabatic holonomies (*3*). To elaborate on how to apply this theory to adiabatic systems, let us revisit the system that is discussed in (*9*). There, adiabatic holonomies with one or two photons are examined in a four-waveguide structure in a tripod arrangement, with couplings that change adiabatically. The cyclic subspace is spanned by time-dependent eigenstates as a cyclic basis. In the single-photon case, there are two-types of eigenmodes that are considered: the two dark modes $\hat{D}_k^\dagger$ $(k = 1,2)$ give rise to eigenstates with eigenenergy zero $\hat{H}\hat{D}_k^\dagger|0\rangle = 0\ \hat{D}_k^\dagger|0\rangle$, while the two bright modes $\hat{B}_\pm^\dagger$ create eigenstates with non-vanishing eigenenergy $\hat{H}\hat{B}_\pm^\dagger|0\rangle = \pm E\ \hat{B}_\pm^\dagger|0\rangle$. In the single-photon case, the dark states $\hat{D}_k^\dagger|0\rangle$ span a degenerate eigenspace that is the cyclic subspace and due to $\langle 0|\hat{D}_k\hat{H}\hat{D}_j^\dagger|0\rangle = 0$ gives rise to a holonomy. Likewise, applying the two-photon holonomic condition from Fig. 2 we find that $\langle 0|\hat{D}_k\hat{D}_j\hat{H}\hat{D}_l^\dagger\hat{D}_m^\dagger|0\rangle = 0$ for all $k, j, l, m = 1,2$. Hence, the subspace that is spanned by the three states of the form $\hat{D}_k^\dagger\hat{D}_j^\dagger|0\rangle$ $(k, j = 1,2$, up to normalization) is a cyclic subspace with vanishing dynamical contribution, giving rise to a two-photon holonomy as discussed in (*9*). However, in this system, $\hat{B}_+^\dagger\hat{B}_-^\dagger|0\rangle$ is another two-photon eigenstate with vanishing eigenvalue. Hence, there is a different holonomic subspace, which is spanned by four eigenstates: three of the form $\hat{D}_k^\dagger\hat{D}_j^\dagger|0\rangle$ $(k, j = 1,2)$ and $\hat{B}_+^\dagger\hat{B}_-^\dagger|0\rangle$. The missing holonomic conditions are

$$\langle 0|\hat{B}_+\hat{B}_-\hat{H}\hat{B}_+^\dagger\hat{B}_-^\dagger|0\rangle = \langle 0|\hat{B}_+\hat{H}\hat{B}_+^\dagger|0\rangle + \langle 0|\hat{B}_-\hat{H}\hat{B}_-^\dagger|0\rangle = E - E = 0$$

and

$$\langle 0|\hat{B}_+\hat{B}_-\hat{H}\hat{D}_k^\dagger\hat{D}_j^\dagger|0\rangle = 0.$$

This larger subspace gives rise to a holonomy, however, as the modes $\hat{D}_k^\dagger$ and $\hat{B}_\pm^\dagger$ do not mix, the state $\hat{B}_+^\dagger\hat{B}_-^\dagger|0\rangle$ does not mix with the other three states either and independently collects a geometric phase contribution.

Generally, if there are degenerate eigenmodes that span a cyclic subspace, as in this example (*9*), this gives rise to an adiabatic single-particle holonomy and consequently a holonomy with larger particle numbers. However, to find an adiabatic two-particle holonomy, it suffices to find degenerate two-particle eigenstates. This approach gives rise to more holonomies and holonomies of larger dimension. In the example (*9*), the mode-based approach shows one two-photon holonomy with three basis states, while the state-based approach additionally finds a two-photon holonomy with four basis states.

Holonomies in the Heisenberg picture

Holonomies and their conditions are usually derived in the Schrödinger picture (*2*, *3*). However, in principal a holonomic condition can also be derived in the Heisenberg picture, which is what we want to do in this section.

Consider a cyclic subspace that is spanned by the propagating modes $\hat{\psi}_k^\dagger$. Let us assume that the $\hat{\psi}_k^\dagger$ do not have an explicit time dependency, i.e. fulfill the Heisenberg equation of motion

$$i d_t \hat{\psi}_k^\dagger = -[\hat{H}, \hat{\psi}_k^\dagger],$$

where $[\hat{H}, \hat{\psi}_k^\dagger] = \hat{H}\hat{\psi}_k^\dagger - \hat{\psi}_k^\dagger\hat{H}$ is the commutator. Let us further assume, that the cyclic subspace has a cyclic basis $\hat{\phi}_k^\dagger$. The following derivation will follow (*2*) closely, however using the Heisenberg picture instead of the Schrödinger picture.

Without loss of generality, let us look at one specific basis state $\hat{\psi}_a^\dagger$ and assume

$$\hat{\psi}_a^\dagger(0) = \hat{\phi}_a^\dagger(0).$$

Then

$$\hat{\psi}_a^\dagger(t) = \sum_b U_{ba}\, \hat{\phi}_b^\dagger(t),$$

where $\hat{U}$ is a unitary matrix. Substituting this into the Heisenberg equation of motion reads

$$\mathrm{id}_t \hat{\psi}_a^\dagger = -[\hat{H}, \hat{\psi}_a^\dagger]$$

$$\sum_b \mathrm{id}_t (U_{ba}\, \hat{\phi}_b^\dagger) = -\sum_b U_{ba}[\hat{H}, \hat{\phi}_b^\dagger]$$

$$\sum_b \mathrm{i}(\mathrm{d}_t U_{ba})\hat{\phi}_b^\dagger + \mathrm{i}U_{ba}(\mathrm{d}_t \hat{\phi}_b^\dagger) = -\sum_b U_{ba}[\hat{H}, \hat{\phi}_b^\dagger].$$

Let us now multiply this equation with the mode $\hat{\phi}_c$ from the left and separately from the right and subtract these two equations.

$$\sum_b \mathrm{i}\, \mathrm{d}_t U_{ba} \left[\hat{\phi}_c, \hat{\phi}_b^\dagger\right] + \mathrm{i}U_{ba}\left[\hat{\phi}_c, \mathrm{d}_t \hat{\phi}_b^\dagger\right] = -\sum_b U_{ba}\left[\hat{\phi}_c, [\hat{H}, \hat{\phi}_b^\dagger]\right]$$

Using the orthogonality of the modes, leads to

$$\mathrm{d}_t U_{ca} = \mathrm{i} \sum_b \mathrm{i}U_{ba}\left[\hat{\phi}_c, \mathrm{d}_t \hat{\phi}_b^\dagger\right] + U_{ba}\left[\hat{\phi}_c, [\hat{H}, \hat{\phi}_b^\dagger]\right],$$

which can be formally integrated as

$$\hat{U} = \mathcal{P} \int_0^T \mathrm{i}\left(\hat{\mathcal{A}} + \hat{\mathcal{K}}\right)\mathrm{d}t$$

with the gauge field in the Heisenberg picture reading $\mathcal{A}_{cb} = \mathrm{i}\left[\hat{\phi}_c, \mathrm{d}_t \hat{\phi}_b^\dagger\right]$ and the dynamical contribution $\mathcal{K}_{cb} = \left[\hat{\phi}_c, [\hat{H}, \hat{\phi}_b^\dagger]\right]$.

Hence, $\left[\hat{\phi}_c, [\hat{H}, \hat{\phi}_b^\dagger]\right] = 0$ for all $c, b$ is the condition for holonomies in the Heisenberg picture. This condition is identical to the one proposed in (*10*) for the special case of linear optical networks. Note, that we did not use any assumption regarding the Hamiltonian $\hat{H}$ apart from hermiticity.

The condition $\left[\hat{\phi}_c, [\hat{H}, \hat{\phi}_b^\dagger]\right] = 0$ will produce holonomies for arbitrary particle numbers. However, it can not be regarded a general holonomic condition. As we show in the main text, any holonomic condition that is based on modes and formulated in the Heisenberg picture will not find all possible holonomies, as not all cyclic subspaces can be described on a mode level.

We therefore argue for addressing the question of holonomies always on a state level, using holonomic conditions from the Schrödinger picture.